\documentclass[]{article}
\usepackage[utf8]{inputenc}

\usepackage{authblk}

\usepackage[sorting=none,sortcites]{biblatex}
\usepackage{amsmath,amssymb,amsfonts,tabu}
\usepackage{physics}
\usepackage{dsfont}
\usepackage{graphicx}
\usepackage{comment}
\usepackage{tikz}
\usepackage{subcaption}

\definecolor{cof}{RGB}{219,144,71}
\definecolor{pur}{RGB}{186,146,162}
\definecolor{greeo}{RGB}{91,173,69}
\definecolor{greet}{RGB}{52,111,72}

\usetikzlibrary{arrows,shapes,trees} 
\usetikzlibrary{positioning,calc}
\usetikzlibrary{decorations.markings}

\newcommand{\Aut}{{\rm Aut}}
\newcommand{\ra}{\rightarrow}
\newcommand{\Hom}{{\rm Hom}}

\newcommand{\Rep}{{\rm Rep}}

\newcommand{\End}{{\rm End}}

\newcommand{\CC}{{\mathbb C}}
\newcommand{\ZZ}{{\mathbb Z}}
\newcommand{\RR}{{\mathbb R}}

\newcommand{\cA}{\mathcal A}
\newcommand{\cB}{\mathcal B}
\newcommand{\cD}{\mathcal D}
\newcommand{\cE}{\mathcal E}
\newcommand{\cL}{\mathcal L}

\newcommand{\cZ}{\mathcal Z}
\newcommand{\cF}{\mathcal F}

\newcommand{\cG}{\mathcal G}

\newcommand{\cK}{{\mathcal K}}
\newcommand{\cJ}{{\mathcal J}}
\newcommand{\cW}{{\mathcal W}}
\newcommand{\cX}{{\mathcal X}}

\newcommand{\AAA}{{\mathbb A}}
\newcommand{\BB}{{\mathbb B}}

\usepackage[left=1in,right=1in,top=1in,bottom=1in]{geometry}

\usepackage{hyperref}
\usepackage{cleveref}

\crefformat{section}{\S#2#1#3}
\crefformat{subsection}{\S#2#1#3}
\crefformat{subsubsection}{\S#2#1#3}

\usepackage{tikz-cd}

\numberwithin{equation}{section}

\hypersetup{colorlinks=true,urlcolor=[rgb]{0,0,0.5},citecolor=[rgb]{0,0.5,0},linkcolor=[rgb]{0,0.5,0}}

\title{Duality and Stacking of Bosonic and Fermionic SPT Phases}
\date{November 30, 2023}

\author[1]{Alex Turzillo\footnote{aturzillo@perimeterinstitute.ca}}
\affil[1]{\small\textit{Perimeter Institute for Theoretical Physics, 
  31 Caroline St N, Waterloo, ON N2L 2Y5, Canada }}

\author[2]{Minyoung You\footnote{miyou849@gmail.com}}
\affil[2]{\small\textit{POSTECH Basic Science Research Institute, Pohang, Gyeongbuk, 37673, Korea}}

\begin{document}

\maketitle

\begin{abstract}

We study the interplay of duality and stacking of bosonic and fermionic symmetry-protected topological phases in one spatial dimension. In general the classifications of bosonic and fermionic phases have different group structures under the operation of stacking, but we argue that they are often isomorphic and give an explicit isomorphism when it exists. This occurs for all unitary symmetry groups and many groups with antiunitary symmetries, which we characterize. We find that this isomorphism is typically not implemented by the Jordan-Wigner transformation, nor is it a consequence of any other duality transformation that falls within the framework of topological holography. Along the way to this conclusion, we recover the fermionic stacking rule in terms of $\cG$-pin partition functions, give a gauge-invariant characterization of the twisted group cohomology invariant, and state a procedure for stacking gapped phases in the formalism of symmetry topological field theory.

\end{abstract}

\tableofcontents

\section{Introduction}

\noindent The celebrated Jordan-Wigner transformation establishes a correspondence between bosonic spin chains and chains of spinless fermions \cite{JordanWigner}. Like that of other duality transformations, this correspondence allows one to reinterpret features of one system in the framework of another. In this paper, we investigate the ability of the Jordan-Wigner transformation and boson-to-fermion dualities in general to translate between key properties of collections of bosonic and fermionic systems -- namely, their behaviors under stacking.

The systems we consider are all one dimensional, gapped, interacting models of spins and fermions with a finite global symmetry $\cG$. These systems are classified into various topological phases characterized by their patterns of breaking and fractionalization of this symmetry. Of particular interest to us are symmetry protected topological (SPT) phases -- those which would be topologically trivial if not for the symmetry. SPT phases of spins (called ``bosonic SPT phases'') and SPT phases of fermions are in one-to-one correspondence under the Jordan-Wigner transformation \cite{FidkowskiKitaev,CGW_complete}. But while the classifications of phases are the same on either side of the duality \emph{as sets}, there is still a sense in which they may differ. Two SPT phases may be ``stacked'' to produce another SPT phase, and this operation gives the collection of SPT phases the structure of an abelian group. The bosonic stacking group law may differ from the fermionic one. The failure of isomorphism of stacking rules is exemplified by systems with $\cG=\ZZ_2^T\times\ZZ_2$ symmetry -- a time-reversal symmetry and a unitary $\ZZ_2$ symmetry that is regarded as fermion parity on the fermionic side of the transformation. The four bosonic SPT phases with this symmetry form the group $\cB=\ZZ_2\times\ZZ_2$ under stacking, while the four fermionic SPT phases (which are modelled by chains of even numbers of Majoranas) stack according to $\cF=\ZZ_4$ \cite{FidkowskiKitaev}. Looking beyond SPT phases for a moment, stacking rules can differ even more drastically: for example, the symmetry-breaking phase of the one dimensional Ising model has no inverse under the stacking operation yet is transformed into an \emph{invertible} topological order by the Jordan-Wigner transformation.

The difference between bosonic and fermionic SPT classification groups has its origins in the bosonic and fermionic commutation relations, so it may be surprising that they ever agree beyond rare accidents. On the other hand, the Jordan-Wigner transformation which relates them is nonlocal in a way that one might expect to compensate for the difference. Our first task is to investigate for which symmetry groups the stacking rules agree and for which they disagree. Our second task is to study whether the Jordan-Wigner transformation or another boson-to-fermion duality can explain this agreement when it occurs. We find that the stacking rules are often isomorphic: they always agree for SPT phases with unitary symmetry (in agreement with a result of Ref. \cite{Kong}), and they may agree in the presence of antiunitary symmetry when certain conditions are satisfied. In the cases where the stacking rules agree, we state an explicit isomorphism between the classification groups and ask whether it is implemented by the Jordan-Wigner transformation or another duality. Typically the answer to this question is ``no.''

The structure of the paper is as follows. In \cref{sec:SPTphases} we review the classifications of bosonic and fermionic SPT phases in terms of group cohomology and supercohomology invariants for general symmetry groups, we define topological and $\cG$-pin partition functions that capture the background-response of these SPT systems, we formulate the Jordan-Wigner transformation in terms of these field theories, and we use them to rederive the SPT stacking rules. In \cref{sec:isomorphic} we derive conditions under which the bosonic and fermionic SPT classifications are isomorphic as groups. These conditions are stated in \cref{sec:necandsuf}. The idea is to understand the classifications as group extensions that become equivalent when a certain extension class vanishes and then to relate the vanishing of this class to the vanishing of certain partition functions. In \cref{sec:dualities}, we study duality and stacking of SPT phases from the point of view of topological holography in order to determine that the isomorphism of classification groups typically cannot be implemented by a physical duality transformation. This result suggests that the stacking rules reflect an essential difference between bosons and fermions that cannot be transformed away by duality and leaves open the mystery of why -- from a physical perspective -- the rules are often isomorphic in the abstract. To reach this conclusion, we understand the action of dualities on the SPT invariants and state a procedure for stacking the Lagrangian algebras that correspond to SPT phases. The appendices contain several related results: we use topological partition functions to rederive the stacking rule for invertible fermionic phases (including the invertible topological order) and review related results in the literature; we use the equivariant state sum construction to evaluate the bosonic partition function on all orientable and nonorientable surfaces, which yields a complete set of gauge-invariant quantities characterizing the SPT invariants; we discuss some interesting computations in twisted group cohomology; and we study how dualities can act on the group of SPT phases, such as by transformations that are not automorphisms.

The authors are grateful for the hospitality of the Asia Pacific Center for Theoretical Physics, the Yukawa Institute for Theoretical Physics, and the Perimeter Institute for Theoretical Physics, where parts of this work were carried out. Research at the Perimeter Institute is supported in part by the Government of Canada through the Department of Innovation, Science and Economic Development and by the Province of Ontario through the Ministry of Colleges and Universities. M.Y. is supported by Basic Science Research Institute Fund, whose NRF grant number is 2021R1A6A1A10042944, the National Research Foundation of Korea (NRF) grant funded by the Korean government (MSIT) (2023R1A2C1006542), and by Grant No. RS-2023-00208291.

\section{Bosonic and fermionic SPT phases}\label{sec:SPTphases}

\subsection{Group cohomology invariants}

The group cohomology invariants of the symmetry group provide a characterization of bosonic SPT phases and -- via the Jordan-Wigner transformation -- fermionic SPT phases as well \cite{FidkowskiKitaev}.
 
Consider bosonic SPT phases. The symmetry class of a bosonic system is specified by a group $\cG$ and a homomorphism $x:\cG\ra\ZZ_2$ that encodes which elements are represented unitarily versus antiunitarily. In this paper, we consider only finite symmetry groups. The classification of bosonic SPT phases with symmetry $(\cG,x)$ is given by classes in the twisted cohomology group \cite{CGLW}
\begin{equation}
    [\omega]\in H^2(\cG,U(1)_x)~.
\end{equation}These are group cochains $\omega:\cG\times\cG\ra U(1)$ satisfying the twisted cocycle condition\footnote{The $U(1)$ coefficients of $\omega$ and the $\ZZ_2$ coefficients of $x$ are written in additive notation, as $U(1)\simeq\RR/\ZZ$ and $\ZZ_2=\{0,1\}$.}
\begin{equation}\label{twistedcocycle}
    0=(\delta\omega)(g,h,k)=(-1)^{x(g)}\omega(h,k)+\omega(g,hk)-\omega(g,h)-\omega(gh,k)
\end{equation}
modulo twisted coboundaries (``gauge transformations'')
\begin{equation}\label{twistedcoboundary}
    (\delta\lambda)(g,h)=(-1)^{x(g)}\lambda(h)+\lambda(g)-\lambda(gh)~.
\end{equation}
The cohomology class $[\omega]$ may be interpreted physically as measuring the projectivity in the representation of the symmetry on the boundary \cite{Pollmann_2010, CGW_complete}. In this paper, we use a partial gauge fixing where $\omega(g,1)=\omega(1,g)=1$, a condition which is preserved by gauge transformations with $\lambda(1)=0$.

Meanwhile, the symmetry class of a fermionic system is specified by bosonic symmetry data $(\cG,x)$ and a distinguished unitary central involution $p\in\cG$ called \emph{fermion parity}. Fermionic SPT phases with symmetry $(\cG,p,x)$ are in one-to-one correspondence with bosonic $(\cG,x)$-SPT phases;\footnote{We do not regard the invertible topological order modeled by the nontrivial Majorana chain as an SPT phase, though it and its symmetry-enrichments are discussed in \cref{sec:invertible}.} namely, they are also associated with classes $[\omega]$. The explicit correspondence between bosonic and fermionic SPT phases given by the Jordan-Wigner transformation will be discussed in detail in \cref{sec:jw}. To characterize certain physical features such as their stacking rule (discussed in \cref{sec:stacking}), however, it is useful to reorganize the topological data into pairs
\begin{equation}\label{grpcohomferm}
    (\omega,\nu)\in C^2(\cG,U(1)_x)\times C^1(\cG,\ZZ_2)
\end{equation}
subject to the constraints\footnote{The map $\tfrac{1}{2}$ embeds $\ZZ_2=\{0,1\}$ as the subgroup $\{0,1/2\}\subset U(1)$.}
\begin{equation}
    \delta\omega=0~,
\end{equation}
\begin{equation}\label{nu}
    \qquad\tfrac{1}{2}\nu(g)=\omega(g,p)-\omega(p,g)+x(g)\omega(p,p)
\end{equation}
modulo coboundaries. Since every closed $\omega$ satisfies the constraint for a unique $\nu$, unchanged by coboundary shift, classes $[\omega,\nu]$ correspond one-to-one with $[\omega]$'s. The invariant $\nu$ singles out the distinguished fermionic parity element\footnote{The invariant $\nu$ is essentially the slant product of $\omega$ with respect to fermion parity $p$.} and encodes whether the boundary action of a symmetry is parity-even or -odd \cite{FidkowskiKitaev}.


\subsection{Supercohomology invariants}\label{sec:supercohomology}

The SPT invariants can be reformulated in the language of supercohomology. This perspective is typically employed to characterize fermionic SPT phases, where the supercohomology invariants describe how a phase arises from a decorated domain wall construction \cite{supercohomology, Wang_2018, Wang_2020}. Similarly, in \cref{sec:fermionicpartitionfunctions} we find supercohomology useful for studying fermionic SPT phases, as it lets us write their explicit partition functions in all symmetry classes. We emphasize, however, that either formulation of the invariants can be used to characterize both bosonic and fermionic SPT phases. As we discuss in \cref{sec:groupexts}, supercohomology has the benefit of putting the bosonic and fermionic SPT stacking rules on an equal footing, as two group extensions of the invariants.

The choice of fermion parity $p$ realizes $\cG$ as a central extension
\begin{equation}
    \ZZ_2^f\ra\cG\xrightarrow[]{b}G_b
\end{equation}
of a group of bosonic symmetries $G_b\simeq\cG/\ZZ_2^f$ by $\ZZ_2^f=\{1,p\}$. It is often useful to describe the symmetry in terms of $G_b$ and a class $[\rho]\in H^2(G_b,\ZZ_2)$ that captures the extension to $\cG$.\footnote{The data $x$ and $\rho$ are sometimes denoted $w_1$ and $w_2$ to reflect that they are Stiefel-Whitney classes of a certain bundle over $BG_b$; however, we avoid this notation in order to distinguish them from the Stiefel-Whitney classes of spacetime below.} Given any function $s:G_b\ra\cG$ with $b\circ s=1$, a cocycle $\rho$ representing the extension class measures the failure of $G_b$ multiplication in $\cG$:
\begin{equation}
    s(g_b)s(h_b)=\rho(g_b,h_b)s(g_bh_b)~,
\end{equation}
and $\rho$ shifts by a coboundary under change of section $s$. We are often interested in symmetries with $[\rho]=0$, in which case $\cG$ \emph{splits} as $G_b\times\ZZ_2$. In this case, a section $s$ with $\rho=0$ is called a splitting.

Using a section $s$, the cochain data on $\cG$ can be pulled back to $G_b$:
\begin{equation}\label{pullbacktoGb}
    \alpha=s^*\omega~,\qquad\beta=s^*\nu~.
\end{equation}
The result is the following data, dubbed \emph{supercohomology classes}: pairs
\begin{equation}
    (\alpha,\beta)\in C^2(G_b,U(1)_x)\times C^1(G_b,\ZZ_2)
\end{equation}
satisfying the conditions
\begin{equation}\label{supercocycle}
    \delta\alpha=\tfrac{1}{2}\beta\cup\rho~,
\end{equation}
\begin{equation}
    \delta\beta=0
\end{equation}
and modulo the relation\footnote{Ref. \cite{TY17}, Theorem A.4 states the relation as $\alpha\mapsto\alpha+\delta\lambda$ for $\lambda:\cG\ra U(1)$ with $\lambda(gp)=\lambda(g)+\lambda(p)$. To get Eq. \eqref{supercoboundary}, note that $(\delta\lambda)(s(g_b),s(h_b))=(-1)^{x(s(g_b))}\lambda(s(h_b))+\lambda(s(g_b))-\lambda(s(g_bh_b))-\lambda(p)\rho(g_b,h_b)$ and take $\lambda_1(g_b)=\lambda(s(g_b))$ and $\lambda_0=\lambda(p)$.}
\begin{equation}\label{supercoboundary}
    \alpha\mapsto \alpha+\delta\lambda_1+\tfrac{1}{2}\lambda_0\cup\rho
\end{equation}
for $\lambda_1:G_b\ra U(1)$ and $\lambda_0\in\ZZ_2$.\footnote{In other words, fermionic SPT phases correspond to cohomology classes of the complex of pairs with differential $\cD(x,y)=(\delta x+\tfrac{1}{2}\rho\cup y,\delta y)$, a.k.a. \emph{supercohomology classes}. Note that $\delta\lambda_0=0$, so $\beta$ is not shifted by a coboundary.} The map \eqref{pullbacktoGb} from group cohomology invariants to supercohomology invariants is bijective, with inverse given on the cochain level\footnote{Every class $[\omega]$ contains a cochain $\omega$ in the partial gauge \eqref{inverseomega}. The remaining freedom in $\omega$ is the relation \eqref{supercoboundary} on $\alpha$.} by
\begin{equation}\label{inverseomega}
    \omega=b^*\alpha+\tfrac{1}{2}b^*\beta\cup t~,\qquad\nu=b^*\beta~,
\end{equation}
where $t$ is the function $t:\cG\ra\ZZ_2^f$ such that $t\circ s=0$ \cite{TY17}. Therefore, supercohomology classes $[\alpha,\beta]$ provide a complete characterization of SPT phases. The set of SPT phases may be expressed as
\begin{equation}\label{supercohomset}
    \AAA\times\BB~,
\end{equation}
where $\AAA$ is the quotient of $H^2(G_b,U(1)_x)$ by the coarser equivalence relation \eqref{supercoboundary}, i.e. classes $[\alpha]$ modulo $[\tfrac{1}{2}\rho]$,\footnote{Technically, due to the condition \eqref{supercocycle}, classes of $\alpha$ belong not to the group $\AAA$ but rather to a torsor over it. If one chooses a ``zero'' $[\alpha_0]$, then each equivalence class $[\alpha]$ may be identified with the element $[\alpha-\alpha_0]$ of $\AAA$.} and $\BB$ is the subgroup of $H^1(G_b,\ZZ_2)$ consisting of \emph{valid} $\beta$ invariants -- that is, cocycles $\beta$ for which there is a solution $\alpha$ to the condition \eqref{supercocycle}, i.e. for which $[\tfrac{1}{2}\beta\cup\rho]=0$.\footnote{The groups $\AAA$ and $\BB$ are sometimes called $H^2_\text{rigid}$ and $BH^1$, and the constraint defining $\BB$ is the twisted Gu-Wen-Freed equation for one spatial dimension. $\AAA$ and $\BB$ are the cokernel and kernel of the spectral sequence differential $[\tfrac{1}{2}\cup\rho]$ \cite{thorngren2019anomalies}.} When $\cG$ splits and $s$ is a splitting, the condition \eqref{supercocycle} and relation \eqref{supercoboundary} reduce to the usual ones in cohomology, and so the set \eqref{supercohomset} becomes\footnote{A third invariant $\gamma\in C^0(G_b,\ZZ_2)=\ZZ_2$ is necessary to classify all invertible fermionic phases with split symmetry group, including symmetry-enrichments of the nontrivial Majorana chain \cite{FidkowskiKitaev,KTY16,Bultinck_2017}. But for SPT phases, $\gamma$ does not appear. The full classification of invertible fermionic phases is discussed in \cref{sec:invertible}.}
\begin{equation}
    H^2(G_b,U(1)_x)\times H^1(G_b,\ZZ_2)~.
\end{equation}
The difference between two sections takes values in $\ZZ_2^f$ and so amounts to a $1$-cochain $\mu:G_b\ra\ZZ_2^f$. Under a change of section, $\rho$ shifts by $\delta\mu$ and $\alpha$ by $\tfrac{1}{2}\beta\cup\mu$ while $\beta$ remains unchanged.\footnote{The choice of section is not physical, yet the classification of phases is also not reduced modulo this shift in $\alpha$; rather, it is only the interpretation of the abstract class $[\alpha,\beta]$ as a physical SPT phase (i.e. as a class $[\omega]$) that depends on the section.} The invariant $\alpha$ represents a $G_b$-symmetric SPT order, while $\beta$ has to do with the decoration of $G_b$ domain walls by complex fermions \cite{supercohomology, Wang_2018, Wang_2020}; the invariant $\beta$ also appears as the fermion parities of twisted sectors \cite{KTY16,KapustinThorgren2017}.


\subsection{Bosonic partition functions}

Now we define topological partition functions parameterized by the group cohomology and supercohomology invariants. These theories represent the responses of SPT phases to background gauge fields for the symmetry and will prove useful later on in \cref{sec:stacking} when we discuss the stacking rule on SPT phases and in \cref{sec:genusrules} when we study the stacking rule isomorphism. The bosonic theory associated to $\omega$ has partition function
\begin{equation}\label{bosonictheory}
    \cZ_{\omega}^b[X,\cA]=\exp(2\pi i\int_X\cA^*\omega)
\end{equation}
on a closed two-dimensional spacetime $X$ with $\cG$-valued background gauge field,\footnote{The map $\cA$ defines a principal $\cG$-bundle on $X$. Since $\cG$ is assumed to be finite, this bundle has a unique (flat) connection.} which is a map $\cA:X\ra B\cG$ satisfying $\cA^*x=[w_1]$, where $[w_i]:X\ra B\ZZ_2$ is the $i^\text{th}$ Stiefel-Whitney class \cite{charclass} of the tangent bundle of $X$ \cite{KT17,kapustincobordism}.\footnote{This constraint enforces that time-reversing holonomies of $\cA$ appear around orientation-reversing cycles of $X$.} The theory is topological due to the condition \eqref{twistedcocycle} and is invariant under the relation \eqref{twistedcoboundary}.

The theory may be reformulated in the supercohomology variables by splitting the $\cG$ gauge field $\cA$ into its $G_b$-valued and $\ZZ_2$-valued parts: a $G_b$ gauge field $A_b:X\ra BG_b$ and a lift of $A_b$ to a $\cG$ gauge field. Following the construction of Dijkgraaf and Witten \cite{DijkgraafWitten},\footnote{On a triangulation of $X$ whose vertices map under $\cA$ to the basepoint of $B\cG$, each edge is assigned an element of $\pi_1(B\cG)=\cG$. Continuity of $\cA$ (mapping triangles to triangles) ensures that the corresponding cochain is closed.} the gauge field $\cA$ is characterized (up to homotopy) by a $\cG$-valued $1$-cochain of $X$ (also denoted by $\cA$) satisfying
\begin{equation}
    \delta\cA=0~,
\end{equation}
\begin{equation}
    \cA^*x=w_1~.
\end{equation}
Meanwhile, the gauge field $A_b$ is represented by a $G_b$-valued cochain satisfying
\begin{equation}\label{Gbgf1}
    \delta  A_b=0~,
\end{equation}
\begin{equation}\label{Gbgf2}
     A_b^*x=w_1~,
\end{equation}
while the lift is characterized by a $\ZZ_2$-valued cochain $A_p$ satisfying\footnote{When $t$ is not closed, it does not induce a map of classifying spaces. As a result, $A_p$ has no corresponding $\ZZ_2$ gauge field.}
\begin{equation}
    \delta A_p=A_b^*\rho~.
\end{equation}
These cochains are related by the correspondence
\begin{equation}
    A_b=\cA^*b~,\qquad A_p=\cA^*t~\qquad\Longleftrightarrow\qquad\cA=A_b^*s+A_p~,
\end{equation}
as can be seen from $\delta t=b^*\rho$. In terms of the pair $(A_b,A_p)$, the bosonic theory \eqref{bosonictheory} becomes\footnote{Each of the two terms in this expression depends on a choice of triangulation of $X$, but their sum does not.}
\begin{equation}
    \cZ_{\alpha,\beta}^b[X,A_b,A_p]=\exp(2\pi i\int_X\cA^*(b^*\alpha+\tfrac{1}{2}b^*\beta\cup t))=\exp(2\pi i\int_XA_b^*\alpha+\tfrac{1}{2}A_b^*\beta\cup A_p)~.
\end{equation}
This partition function is section-independent by construction, but one can also verify that the shift $\frac{1}{2}\beta\cup\mu$ in $\alpha$ under change of section is cancelled exactly by the shift $A_b^*\mu$ in $A_p$.

\subsection{Fermionic partition functions}\label{sec:fermionicpartitionfunctions}

Fermionic partition functions are sensitive to $\cG$-pin structures rather than to $\cG$ gauge fields. A $\cG$-pin structure $\eta^\cG$ can -- like the gauge field $\cA$ -- be formulated as a $\cG$-valued $1$-cochain, this time satisfying\footnote{The first condition means that the violation of the triple overlap condition by $\eta^\cG$ is valued in $\ZZ_2$ and coincides with that of a pin-minus structure. Compared to the treatment of $\cG$-spin structures in Ref. \cite{KTY16}, we have absorbed the freedom $\phi$. See Ref. \cite{thorngren2019anomalies} for another perspective on $\cG$-pin structures, referred to as twisted spin structures.}
\begin{equation}
    \delta\eta^\cG=w_1^2+w_2~,
\end{equation}
\begin{equation}
    \eta^{\cG*}x=w_1~.
\end{equation}
It can be re-expressed as a $G_b$ gauge field $A_b$ satisfying \eqref{Gbgf1} and \eqref{Gbgf2} and a $\ZZ_2$-valued cochain $\eta$ satisfying
\begin{equation}
    \delta\eta=w_1^2+w_2+A_b^*\rho
\end{equation}via the correspondence
\begin{equation}
    A_b=\eta^{\cG*}b~,\qquad\eta=\eta^{\cG*}t\qquad\Longleftrightarrow\qquad\eta^\cG=A_b^*s+\eta~.
\end{equation}
$\cG$-pin structures generalize many familiar structures. A $\cG$-spin structure is realized when all symmetries are unitary ($x=0$), in which case the theory is defined only on orientable surfaces ($w_1=0$). When $\cG$ splits and $s$ is a splitting, the map $\eta$ is a pin-minus structure (trivialization of $w_1^2+w_2$); in particular, for $\cG=\ZZ_2^T\times\ZZ_2^f$, $A_b$ is completely fixed by $w_1$, and so a $\cG$-pin structure is just the pin-minus structure $\eta$. A pin-plus structure (trivialization of $w_2$) is realized for $\cG=\ZZ_4^{Tf}$ since $A_b$ is completely fixed by $w_1$ and $\rho=x\cup x$ means $\delta\eta=w_2$.

To write the fermionic partition functions, we make use of the quadratic enhancement of the cup pairing that is associated to every pin-minus structure $e$ \cite{KT89,atiyahspin,johnsonspin}. This is a map $q_e:H^1(X,\ZZ_2)\ra\ZZ_4$ satisfying
\begin{equation}\label{quadratic}
    q_e(a)+q_e(b)-q_e(a+b)=2\int_X a\cup b~,
\end{equation}
\begin{equation}\label{shiftrule}
    q_{e+b}(a)-q_e(a)=2\int_X a\cup b
\end{equation}
for $\ZZ_2$ gauge fields $a$ and $b$. To work with $\cG$-pin structures, we introduce an auxiliary $\ZZ_2$-valued cochain $\tau$ satisfying $\delta\tau=A_b^*\rho$, so that $e=\eta+\tau$ is a pin-minus structure. Given the existence of a $\cG$-pin structure on $X$, a solution $\tau$ always exists because every two-dimensional space is pin-minus \cite{KT89}. The partition function should not depend on the choice of $\tau$. In terms of the pair $(A_b,\eta)$ and the auxiliary variable $\tau$, the theory is
\begin{equation}\label{fermionictheory}
    \cZ_{\alpha,\beta}^f[X,A_b,\eta]=\exp(2\pi i\int_XA_b^*\alpha+\tfrac{1}{2}A_b^*\beta\cup\tau)\exp(\frac{\pi i}{2}q_{\eta+\tau}(A_b^*\beta))~,
\end{equation}
Note that our expression \eqref{fermionictheory} reduces to the $\cG$-spin partition function of Ref. \cite{KTY16} in the case of unitary symmetries, where $q/2$ defines a quadratic refinement associated to a spin structure since $q$ is even on orientable manifolds.  To see that the theory \eqref{fermionictheory} is well-defined, several properties must be checked. First, the theory is independent of the auxiliary variable $\tau$ since shifting $\tau$ by a closed $b$ changes each of the two $\tau$-dependent terms by the same sign $A_b^*\beta\cup b$, as follows from the relation \eqref{shiftrule}. Second, the theory is independent of the section $s$ since $\alpha$ shifts by $\tfrac{1}{2}\beta\cup\mu$ and $\eta$ shifts by $A_b^*\mu$ and so by the same rule \eqref{shiftrule} these variations cancel. Third, the theory is topological since $\delta(A_b^*\alpha+\tfrac{1}{2}A_b^*\beta\cup\tau)=0$ by the conditions \eqref{supercocycle}. Fourth, the theory is invariant under the relation \eqref{supercoboundary} since $b^*(\lambda_0\cup\rho)=\delta(b^*\lambda_0\cup t)$. 

In the split case, the auxiliary variable may be eliminated by setting it to $\tau=0$. Then the theory becomes
\begin{equation}\label{fermionictheorysplit}
    \cZ_{\alpha,\beta}^f[X,A_b,\eta]=\exp(2\pi i\int_XA_b^*\alpha)\exp(\frac{\pi i}{2}q_\eta(A_b^*\beta))~.
\end{equation}

\subsection{The Jordan-Wigner transformation}\label{sec:jw}

Now we are ready to define the Jordan-Wigner transformation $\cJ\cW$ on partition functions. It maps a bosonic partition function to the fermionic partition function defined as follows:
\begin{equation}\label{jordanwigner}
    \cJ\cW(\cZ^b)[X,A_b,\eta]:=\sum_{A_p}\frac{\text{Arf}(\eta+A_p)}{\sqrt{|H^1(X,\ZZ_2)|}}\cZ^b[X,A_b,A_p]~.
\end{equation}
Here we have used the Arf-Brown-Kervaire invariant \cite{Kervaire}
\begin{equation}\label{arf}
    \text{Arf}(e):=\frac{1}{\sqrt{|H^1(X,\ZZ_2)|}}\sum_{a\in H^1(X,\ZZ_2)}\exp(\frac{\pi i}{2}q_e(a))
\end{equation}
of a pin-minus structure $e$. Some useful properties of the Arf invariant are collected in \cref{sec:arf}. Using the property \eqref{qtimesarf}, we can introduce an auxiliary variable $\tau$ to rewrite the transformation as
\begin{equation}\label{jordanwignertau}
    \cJ\cW(\cZ^b)[X,A_b,\eta]=\frac{\text{Arf}(\eta+\tau)}{\sqrt{|H^1(X,\ZZ_2)|}}\sum_{A_p}\cZ^b[X,A_b,A_p]\exp(\frac{-\pi i}{2}q_{\eta+\tau}(A_p+\tau))~.
\end{equation}
As demonstrated in \cref{sec:arf}, the inverse of the Jordan-Wigner transformation is given by
\begin{equation}\label{jwinverse}
    \cJ\cW^{-1}(\cZ^f)[X,A_b,A_p]=\sum_\eta\frac{\text{Arf}(\eta+A_p)^{-1}}{\sqrt{|H^1(X,\ZZ_2)|}}\cZ^f[X,A_b,\eta]~.
\end{equation}
These forms of the Jordan-Wigner transformation were first stated in Ref. \cite{GK} for split, unitary symmetries and were generalized to include antiunitary symmetries in Ref. \cite{thorngren2019anomalies}.

The Jordan-Wigner transform of the bosonic theory associated with the invariant $[\alpha,\beta]$ is the fermionic theory associated with the same invariant, as can be seen by computing
\begin{align}\begin{split}
    \cJ\cW(\cZ^b_{\alpha,\beta})[X,A_b,\eta]
    &=\frac{\text{Arf}(\eta+\tau)}{\sqrt{|H^1(X,\ZZ_2)|}}\sum_{A_p}\exp(2\pi i\int_XA_b^*\alpha+\tfrac{1}{2}A_b^*\beta\cup A_p)\exp(\frac{-\pi i}{2}q_{\eta+\tau}(A_p+\tau))\\
    &=\exp(2\pi i\int_XA_b^*\alpha+\tfrac{1}{2}A_b^*\beta\cup\tau)\exp(\frac{\pi i}{2}q_{\eta+\tau}(A_b^*\beta))\\
    &\qquad\qquad\qquad\qquad\times\,\frac{\text{Arf}(\eta+\tau)}{\sqrt{|H^1(X,\ZZ_2)|}}\sum_{A_p}\exp(\frac{-\pi i}{2}q_{\eta+\tau}(A_b^*\beta+A_p+\tau))\\
    &=\cZ^f_{\alpha,\beta}[X,A_b,\eta]~,
\end{split}\end{align}
where the second line follows from the relation \eqref{quadratic} and the third from the expression \eqref{invarf}. This means that the Jordan-Wigner transformation defines a map from bosonic to fermionic SPT phases given by
\begin{equation}\label{jwsupcohom}
    \cJ\cW:[\alpha,\beta]\mapsto[\alpha,\beta]~.
\end{equation}

\subsection{Stacking rules}\label{sec:stacking}

As we have just observed, the sets of bosonic and fermionic SPT phases are equal, with the Jordan-Wigner transformation defining a bijective correspondence between them. However, there is a natural group operation defined on the two sets given by \emph{stacking}, and the classifications are \emph{not} always isomorphic as groups. A simple example is given by SPT phases with symmetry $\cG=\ZZ_2^T\times\ZZ_2^f$: there are four SPT phases on each side of the correspondence; however, the bosonic phases form the group $\cB=\ZZ_2\times\ZZ_2$ under stacking while the fermionic phases form the group $\cF=\ZZ_4$ \cite{FidkowskiKitaev}. This and other examples are explored in \cref{sec:examples}.

The \emph{stack} of two theories is the theory whose partition function is the product.\footnote{In the language of lattice models, the stack of two systems is the one defined on the tensor product Hilbert space with tensor product time evolution $U_{AB}=U_A\otimes U_B$. This means that the stacked Hamiltonian is $H_{AB}=H_A\otimes\mathds{1}+\mathds{1}\otimes H_B$.} From the expression \eqref{bosonictheory} for the bosonic partition function in the group cohomology variable, it follows that
\begin{equation}\label{bosonicomegapartitionfunctionstacking}
    \cZ^b_{\omega_1}\cZ^b_{\omega_2}=\cZ^b_{\omega_1+\omega_2}~,
\end{equation}
so the invariants stack according to the usual group law on $H^2(\cG,U(1)_x)$:
\begin{equation}\label{bosonicstackingomega}
    [\omega_1]\otimes_\cB[\omega_2]=[\omega_1+\omega_2]~.
\end{equation}
On the other hand, in the variables \eqref{grpcohomferm}, the fermionic stacking rule is given by
\begin{equation}
    [\omega_1]\otimes_\cF[\omega_2]=[\omega_1+\omega_2+\tfrac{1}{2}\nu_1\nu_2]~.
\end{equation}
In terms of the supercohomology invariants, these stacking rules read
\begin{equation}\label{bosonicstacking}
    \left[\begin{array}{c}\alpha_1\\\beta_1\end{array}\right]\otimes_\cB\left[\begin{array}{c}\alpha_2\\\beta_2\end{array}\right]=\left[\begin{array}{c}\alpha_1+\alpha_2\\\beta_1+\beta_2\end{array}\right]
\end{equation}
and
\begin{equation}\label{fermionicstacking}
    \left[\begin{array}{c}\alpha_1\\\beta_1\end{array}\right]\otimes_\cF\left[\begin{array}{c}\alpha_2\\\beta_2\end{array}\right]=\left[\begin{array}{c}\alpha_1+\alpha_2+\tfrac{1}{2}\beta_1\beta_2\\\beta_1+\beta_2\end{array}\right]~.
\end{equation}
The supercohomology formulation of the fermionic stacking rule was first stated in Ref. \cite{GK} in the case of a split, unitary symmetry group. It was generalized to arbitrary unitary symmetries in Refs. \cite{Bultinck_2017,KTY16} and to the most general setting in Ref. \cite{Bultinck_2017}. The general rule has been reproduced in Refs. \cite{TY17,Bourne,Aksoy}.

When the symmetry is unitary and split, the group of classes $[\alpha,\beta]\in H^2(G_b,U(1)_x)\times H^1(G_b,\ZZ_2)$ under the fermionic stacking rule \eqref{fermionicstacking} is isomorphic to $\tilde\Omega_\text{spin}^2(BG_b)$, the reduced spin cobordism group\footnote{The ``cobordism'' group refers to the Pontryagin dual $\Hom(\,\cdot\,,U(1))$ of the bordism group. The bordism group that appears is reduced because we have specialized to SPT phases among all invertible fermionic phases.} of $BG_b$ \cite{kapustincobordism,kttw,GK,pontryagindual}. More generally, the group of fermionic SPT phases is given by \emph{twisted} cobordism \cite{kttw,thorngren2019anomalies}:
\begin{equation}
    \cF(G_b,\rho,x)\simeq\tilde\Omega_\text{spin}^2(BG_b,\rho,x)~.
\end{equation}
This is in contrast with bosonic SPT phases, whose group structure \eqref{bosonicstacking} is that of twisted cohomology
\begin{equation}
    \cB(G_b,\rho,x)\simeq H^2(\cG,U(1)_x)~.
\end{equation}

The bosonic and fermionic stacking rules can be recovered by multiplying partition functions. From
\begin{align}\begin{split}
    \cZ_{\alpha_1,\beta_1}^b[X,A_b,A_p]\cZ_{\alpha_2,\beta_2}^b[X,A_b,A_p]
    &=\exp(2\pi i\int_XA_b^*(\alpha_1+\alpha_2)+\tfrac{1}{2}A_b^*(\beta_1+\beta_2)\cup A_p)\\
    &=\cZ_{\alpha_1+\alpha_2,\beta_1+\beta_2}^b[X,A_b,A_p]~,
\end{split}\end{align}
one sees the rule \eqref{bosonicstacking}. And due to the quadratic relation \eqref{quadratic}, one sees the fermionic rule \eqref{fermionicstacking} from
\begin{align}\begin{split}
    \cZ_{\alpha_1,\beta_1}^f[X,A_b,\eta]&\cZ_{\alpha_2,\beta_2}^f[X,A_b,\eta]\\
    &=\exp(2\pi i\int_XA_b^*(\alpha_1+\alpha_2)+\tfrac{1}{2}A_b^*(\beta_1+\beta_2)\cup\tau)\exp(\frac{\pi i}{2}(q_{\eta+\tau}(A_b^*\beta_1)+q_{\eta+\tau}(A_b^*\beta_2)))\\
    &=\exp(2\pi i\int_XA_b^*(\alpha_1+\alpha_2+\tfrac{1}{2}\beta_1\cup\beta_2)+\tfrac{1}{2}A_b^*(\beta_1+\beta_2)\cup\tau)\exp(\frac{\pi i}{2}q_{\eta+\tau}(A_b^*(\beta_1+\beta_2)))\\
    &=\cZ_{\alpha_1+\alpha_2+\tfrac{1}{2}\beta_1\cup\beta_2,\beta_1+\beta_2}^f[X,A_b,\eta]~.
\end{split}\end{align}
In \cref{sec:invertible}, this approach of multiplying partition functions is used to extend the fermionic stacking rule to all invertible phases, including those beyond SPT phases.

\section{Isomorphic classifications}\label{sec:isomorphic}

\subsection{The classifications as group extensions}\label{sec:groupexts}

We now ask the following: when are the bosonic and fermionic SPT classifications isomorphic as groups? We find that, despite their different looking forms \eqref{bosonicstacking} and \eqref{fermionicstacking} in the supercohomology variables, the groups are often isomorphic, even when the Jordan-Wigner transformation does not provide an isomorphism.

Each of the two stacking rules $\cB$ and $\cF$ can be realized as a group extension of $\BB$ by $\AAA$:
\begin{equation}
    \AAA\longrightarrow\cB\longrightarrow\BB~,\qquad\AAA\longrightarrow\cF\longrightarrow\BB~,
\end{equation}
where $\AAA$ and $\BB$ were first defined in \cref{sec:supercohomology} as the following: $\AAA$ is the quotient of $H^2(G_b,U(1)_x)$ by $\ZZ_2=\langle\tfrac{1}{2}\rho\rangle$, and $\BB$ is the subgroup of $H^1(G_b,\ZZ_2)$ consisting of \emph{valid} $\beta$, i.e. those satisfying $[\tfrac{1}{2}\beta\cup\rho]=0$ (equivalently, those that pullback from $\nu$ \eqref{nu} associated with $\omega$). Since the extensions are central (as $\cB$ and $\cF$ are abelian by construction), each is characterized by an extension class in $[\Omega_{\cB,\cF}]\in H^2(\BB,\AAA)$ that appears as
\begin{equation}\label{bothstacking}
    \left[\begin{array}{c}\alpha_1\\\beta_1\end{array}\right]\otimes_{\cB,\cF}\left[\begin{array}{c}\alpha_2\\\beta_2\end{array}\right]=\left[\begin{array}{c}\alpha_1+\alpha_2+\Omega_{\cB,\cF}(\beta_1,\beta_2)\\\beta_1+\beta_2\end{array}\right]~.
\end{equation}
The expressions \eqref{bosonicstacking} and \eqref{fermionicstacking} reveal that
\begin{equation}
    \Omega_\cB=0~,\qquad\Omega_\cF=\tfrac{1}{2}\cup~.
\end{equation}
The bosonic classification group is simply the direct product group\footnote{Since, due to the condition \eqref{supercocycle}, a equivalence class of an $\alpha$ is not canonically identified with an element of $\AAA$ (but rather to an element of a torsor over it), this isomorphism is also noncanonical.}
\begin{equation}
    \cB\simeq\AAA\times\BB~,
\end{equation}
and we are interested in studying when $\cF\simeq\cB\simeq\AAA\times\BB$. Since the direct product group can arise only from the trivial extension \cite{Ayoub}, this only happens when $[\Omega_\cF]=0$, i.e. when there exists a $\Lambda:\BB\ra\AAA$ such that
\begin{equation}\label{trivialextension}
    \Omega_\cF=\delta\Lambda~.
\end{equation}
In this case, the map
\begin{equation}\label{isommap}
    \cX:[\alpha,\beta]\mapsto[\alpha-\Lambda(\beta),\beta]
\end{equation}
gives an isomorphism of stacking group laws:
\begin{align}\begin{split}
    \cX[\alpha_1,\beta_1]\otimes_\cF\cX[\alpha_2,\beta_2]
    &=[\alpha_1-\Lambda(\beta_1),\beta_1]\otimes_\cF[\alpha_2-\Lambda(\beta_2),\beta_2]\\
    &=[\alpha_1+\alpha_2-\Lambda(\beta_1)-\Lambda(\beta_2)+\Omega_\cF(\beta_1,\beta_2),\beta_1+\beta_2]\\
    &=[\alpha_1+\alpha_2-\Lambda(\beta_1+\beta_2),\beta_1+\beta_2]\\
    &=\cX[\alpha_1+\alpha_2,\beta_1+\beta_2]\\
    &=\cX([\alpha_1,\beta_1]\otimes_\cB[\alpha_2,\beta_2])~.
\end{split}\end{align}
When the isomorphism $\cX$ exists, it is in general not unique, as $\Lambda$ may be shifted by any cocycle in $Z^1(\BB,\AAA)$.

We emphasize that the condition \eqref{trivialextension} is weaker than asking when $\Omega_\cF(\beta_1,\beta_2)$ vanishes in $\AAA$ (that is, when $\tfrac{1}{2}\beta_1\cup\beta_2$ is cohomologous to $\tfrac{1}{2}\rho$) for all $\beta_1,\beta_2\in\BB$. The vanishing of $\Omega_\cF$ for all inputs means that the Jordan-Wigner transformation \eqref{jwsupcohom} is an isomorphism, whereas the condition \eqref{trivialextension} only means that there is \emph{some} isomorphism $\cX:\cB\ra\cF$, not necessarily given by $\cJ\cW$. We illustrate this subtlety with some simple examples in \cref{sec:examples}, and we ask about the physical meaning of these other isomorphisms in \cref{sec:dualities}.

Let us briefly comment on the question of when $\cJ\cW$ is an isomorphism of stacking rules, in the simple case where the symmetry $\cG$ is finite abelian and unitary. The group $\cG$ is a product of cyclic groups $\ZZ_{n_i}$, and (up to isomorphism) $p$ belongs to one of them, $\ZZ_{n_f}^f$, of order $n_f$ a power of $2$, as the element $p=n_f/2$. Let $n_{ij}$ denote $\gcd(n_i,n_j)$. We claim that $\cJ\cW$ is an isomorphism precisely when there is at most one factor $\ZZ_{n_i}$, $i\ne f$, with $n_{if}=n_f$, i.e. where $n_i$ contains at least as many factors of $2$ than $n_f$.\footnote{This notion is well-defined because factors of even order cannot be combined since their orders are not coprime.} The group cohomology of $\cG$ is generated by cocycles $\omega_{ij}(g,h)=\tfrac{1}{n_{ij}}g_ih_j$. Valid $\beta$ are generated by $\beta_i(g)=\omega_{if}(g,p)=\frac{n_f}{2n_{if}}g_i=\tfrac{1}{2}g_i$ for $i$ with $n_{if}=n_f$. If there is at most one factor with $n_{if}=n_f$, there is at most one nonzero valid $\beta$, so $\Omega_\cF$ vanishes and $\cJ\cW$ is an isomorphism. When there is more than one such factor, there are valid $\beta_i,\beta_j$, each evaluating to $1$ on distinct cyclic generators $g,h$ and $0$ on the others. Then $\Omega_\cF(\beta_i,\beta_j)=\tfrac{1}{2}\beta_i\cup\beta_j=\tfrac{n_{ij}}{2}\omega_{ij}$, which is nontrivial in $\AAA$, so $\cJ\cW$ is not an isomorphism. Further examples are discussed in \cref{sec:examples}.

\subsection{Examples}\label{sec:examples}

Before answering the question in general, let us consider a few examples.

First is $\cG=\ZZ_2^T\times\ZZ_2^f$, where there is no isomorphism. There are four SPT phases, each characterized by a choice of supercohomology data -- one of the two possible $\alpha$ and one of the two possible $\beta$:
\begin{equation}
    \alpha_0(t,t)=0~,\qquad\alpha_1(t,t)=\tfrac{1}{2}~,
\end{equation}
\begin{equation}
    \beta_0(t)=0~,\qquad\beta_1(t)=1~.
\end{equation}
These cocycles add according to
\begin{equation}\label{addexample}
    \alpha_1+\alpha_1=\alpha_0~,\qquad\beta_1+\beta_1=\beta_0~,
\end{equation}
so, denoting each cocycle by its subscript, we have the bosonic stacking rule
\begin{equation}
    \left[\begin{array}{c}i\\j\end{array}\right]\otimes_\cB\left[\begin{array}{c}i'\\j'\end{array}\right]=\left[\begin{array}{c}i+i'\\j+j'\end{array}\right]~,
\end{equation}
which is the classification group $\cB=\ZZ_2\times\ZZ_2$. The only nontrivial cup product of $\beta$ is
\begin{equation}
    \tfrac{1}{2}\beta_1\cup\beta_1=\alpha_1~,
\end{equation}
so the fermionic stacking rule is
\begin{equation}
    \left[\begin{array}{c}i\\j\end{array}\right]\otimes_\cF\left[\begin{array}{c}i'\\j'\end{array}\right]=\left[\begin{array}{c}i+i'+jj'\\j+j'\end{array}\right]~,
\end{equation}
which is the classification group $\cF=\ZZ_4$, generated by the phase $[\alpha_0,\beta_1]$. Therefore $\cB\not\simeq\cF$.

A second example is $\cG=\ZZ_4^T\times\ZZ_2^f$, where now $t$ is of order four. Here, there is an isomorphism given by the Jordan-Wigner transformation. Again there are two each of $\alpha$ and $\beta$:
\begin{equation}
    \alpha_0(t^m,t^n)=0~,\qquad\alpha_1(t^m,t^n)=\tfrac{1}{4}[m]_2n~,
\end{equation}
\begin{equation}
    \beta_0(t^m)=0~,\qquad\beta_1(t^m)=[m]_2,
\end{equation}
where $[\,\cdot\,]_2$ denotes reduction mod $2$. The cochain $\alpha_1$ is indeed a twisted cocycle since
\begin{align}\begin{split}
    (\delta\alpha)_1(t^l,t^m,t^n)
    &=(-1)^l\tfrac{1}{4}[m]_2n+\tfrac{1}{4}[l]_2(m+n)-\tfrac{1}{4}[l]_2m-\tfrac{1}{4}[l+m]_2n\\
    &=\tfrac{1}{4}n([m]_2-2[l]_2[m]_2+[l]_2-[l+m]_2)\\
    &=0
\end{split}\end{align}
and is nontrivial since
\begin{equation}
    \alpha_1(t,t^3)-\alpha_1(t^3,t)=\tfrac{1}{2}
\end{equation}
is a nonzero gauge invariant quantity. The cocycles add as in the previous example \eqref{addexample}, so the bosonic classification is again $\cB=\ZZ_2\times\ZZ_2$. But this time, the cup products are all trivial in cohomology:
\begin{equation}
    (\tfrac{1}{2}\beta_1\cup\beta_1)(t^m,t^n)=\tfrac{1}{2}[m]_2[n]_2=\tfrac{1}{4}((m-[m]_2)+(-1)^m(n-[n]_2)-(m+n-[m+n]_2))~,
\end{equation}
\begin{equation}
    \Longrightarrow\qquad\tfrac{1}{2}\beta_1\cup\beta_1=\delta\lambda~,\qquad\lambda(t^m)=\tfrac{1}{4}(m-[m]_2)~,
\end{equation}
so the fermionic stacking rule is also $\cF=\ZZ_2\times\ZZ_2$. In this example, the Jordan-Wigner transformation is an isomorphism $\cB\ra\cF$ since the bosonic and fermionic stacking rules are identical in terms of $\alpha_i,\beta_i$.

Finally, consider the symmetry group $\cG=\ZZ_2\times\ZZ_2\times\ZZ_2^f$, where there is an isomorphism $\cB\simeq\cF$ but it is not given by the Jordan-Wigner transformation. This group has two $\alpha$ and four $\beta$:
\begin{equation}
    \alpha_0((a,b),(a',b'))=0~,\qquad\alpha_1((a,b),(a',b'))=\tfrac{1}{2}a'b~,
\end{equation}
\begin{equation}
    \beta_{00}(a,b)=0~,\qquad\beta_{10}(a,b)=a~,\qquad\beta_{01}(a,b)=b~,\qquad\beta_{11}(a,b)=[a+b]_2~,
\end{equation}
written in terms of elements $(a,b)\in\ZZ_2\times\ZZ_2=G_b$. These cocycles add according to
\begin{equation}
    \alpha_i+\alpha_{i'}=\alpha_{i+i'}~,\qquad\beta_{j,k}+\beta_{j',k'}=\beta_{(j+j'),(k+k')}~,
\end{equation}
so the bosonic stacking rule is $\cB=\ZZ_2^3$:
\begin{equation}
    \left[\begin{array}{c}i\\j\\k\end{array}\right]\otimes_\cB\left[\begin{array}{c}i'\\j'\\k'\end{array}\right]=\left[\begin{array}{c}i+i'\\j+j'\\k+k'\end{array}\right]~,
\end{equation}
where the second and third rows represent $\beta_{10}$ and $\beta_{01}$. The cup products are
\begin{equation}
    (\tfrac{1}{2}\beta_{jk}\cup\beta_{j'k'})((a,b),(a',b'))=\tfrac{1}{2}(ja+kb)(j'a'+k'b')=\tfrac{1}{2}jj'aa'+\tfrac{1}{2}jk'ab'+\tfrac{1}{2}kj'ba'+\tfrac{1}{2}kk'bb'
\end{equation}
\begin{equation}
    \Longrightarrow\qquad\tfrac{1}{2}\beta_{jk}\cup\beta_{j'k'}\sim(jk'+j'k)\,\alpha_1
\end{equation}
since, in cohomology, $\tfrac{1}{2}aa'$ and $\tfrac{1}{2}bb'$ are trivial and $\tfrac{1}{2}ab'\sim\tfrac{1}{2}a'b$. Therefore the fermionic stacking rule is
\begin{equation}
    \left[\begin{array}{c}i\\j\\k\end{array}\right]\otimes_\cF\left[\begin{array}{c}i'\\j'\\k'\end{array}\right]=\left[\begin{array}{c}i+i'+jk'+j'k\\j+j'\\k+k'\end{array}\right]~.
\end{equation}
Because the bosonic and fermionic stacking rules are not identical, the Jordan-Wigner transformation is not an isomorphism. However, a trivialization of the extension class is given by
\begin{equation}
    \Lambda(\beta_{jk})=jk\,\alpha_1~,\qquad\text{i.e.}\qquad\Lambda(\beta_{11})=\alpha_1~,\quad\Lambda(\text{else})=0~,
\end{equation}
which has
\begin{equation}
    (\delta\Lambda)(\beta,\beta')=jk\,\alpha_1+j'k'\,\alpha_1-(j+j')(k+k')\,\alpha_1=-(jk'+j'k)\,\alpha_1~.
\end{equation}
This trivialization is a simple case of the more general form \eqref{lambdaexplicit}. Therefore the map
\begin{equation}
    \cX:\left[\begin{array}{c}i\\j\\k\end{array}\right]\mapsto\left[\begin{array}{c}i+jk\\j\\k\end{array}\right]
\end{equation}
from bosonic to fermionic SPT phases is an isomorphism, and the fermionic stacking rule is $\cF=\ZZ_2^3$. We will return to this example in \cref{dualities:example} when we study whether the map $\cX$ can be implemented by a duality.

\subsection{Necessary and sufficient conditions for isomorphism}\label{sec:necandsuf}

In order to state the necessary and sufficient conditions for the bosonic and fermionic SPT classifications to be isomorphic, we first define several conditions. We call these conditions ``genus rules'' in reference to the fact, explained in \cref{sec:genusrules}, that the genus $n$ rule is equivalent to the vanishing of the bosonic partition function associated to the cocycles $\omega=\tfrac{1}{2}b^*\beta\cup b^*\beta$, for all valid $\beta$, on the nonorientable surface of genus $n$.

\noindent\hspace{0.03\textwidth}\fbox{\label{genusrule}
\hspace{0.01\textwidth}\begin{minipage}{0.90\textwidth}
\vspace{1mm}
\noindent\emph{Genus $n$ rule.}
\vspace{1mm}

A fermionic symmetry group $(G_b,\rho,x)$ is said to satisfy the genus $n$ rule if
\begin{equation}\label{genusruleeq}
    \sum_i^nb^*\beta(g_i)=0\text{ mod }2~.
\end{equation}
for every $\beta:G_b\ra\ZZ_2$ with $[\tfrac{1}{2}\beta\cup\rho]=0$ and elements $g_1,\ldots,g_n\in\cG$ with $x(g_i)=1$ and $\prod_i^ng_i^2=1$. 
\vspace{-2mm}
\end{minipage}
\hspace{0.01\textwidth}
}

\noindent In the split case,\footnote{If the group does not split, one must worry about $G_b$ gauge fields $A_b=\{g_{b1},\ldots,g_{bn}\}$ that do not extend to $\cG$ gauge fields. These have $\int_X A_b^*\rho\ne 0$ (and no solution $A_p$), so they do not distinguish when a class is trivial in $\AAA$.} the genus $n$ rule may be framed in terms of $G_b$ elements as the condition that
\begin{equation}
    \sum_i^n\beta(g_{bi})=0\text{ mod }2=0
\end{equation}
for every $\beta:G_b\ra\ZZ_2$ and any elements $g_{b1},\ldots,g_{bn}\in G_b$ with $x(g_{bi})=1$ and $\prod_i^ng_{bi}^2=1$.

If the genus $n$ rule is satisfied for some $n>2$, then it is also satisfied for $n-2$. To see this, suppose that the genus $n-2$ rule is violated, which means that there exists valid $\beta$ and antiunitary $g_1,\ldots,g_{n-2}$ satisfying $\prod_i^{n-2}g_i^2=1$ but with $\sum_i^{n-2}b^*\beta(g_i)=1$. Then, for any antiunitary $g$, taking $g_{n-1}=g$ and $g_n=g^{-1}$ means that the sequence $g_1,\ldots,g_n$ satisfies $\prod_i^ng_i^2=1$ but has $\sum_i^nb^*\beta(g_i)=1$, violating the genus $n$ rule.

When the symmetry group $\cG$ is abelian, the relation also goes the other way: satisfying the genus $n$ rule implies satisfying the genus $n+2$ rule. To see this, suppose that the genus $n+2$ rule is violated, which means that there exists valid $\beta$ and antiunitary $g_1,\cdots,g_{n+2}$ with $\prod_i^{n+2}g_i^2=1$ and $\sum_i^{n+2}b^*\beta(g_i)=1$. Then define $g=g_ng_{n+1}g_{n+2}$, which is antiunitary, has $\sum_i^{n-1}b^*\beta(g_i)+b^*\beta(g)=1$, and, since the group is abelian, satisfies $(\prod_i^{n-1}g_i^2)g^2=1$. Therefore $\cG$ violates the genus $n$ rule with the sequence $g_1,\ldots,g_{n-1},g$.

A group is said to satisfy the \emph{genus $\infty$ rule} if it satisfies the genus $n$ rule for every $n$. For abelian groups, the genus $\infty$ rule is equivalent to the genus $1$ and $2$ rules together, by the argument above.

In terms of these rules, the following necessary and sufficient conditions for an isomorphism $\cB\simeq\cF$ hold.

\noindent\hspace{0.03\textwidth}\fbox{
\hspace{0.01\textwidth}\begin{minipage}{0.90\textwidth}
\vspace{1mm}
\noindent\emph{General result.}
\vspace{1mm}

For the groups of bosonic and fermionic SPT phases to be isomorphic
\begin{equation}
    \cB(G_b,\rho,x)\simeq\cF(G_b,\rho,x)~,
\end{equation}
\begin{itemize}
    \item The genus $1$ rule is a necessary condition.
    \item The genus $\infty$ rule is a sufficient condition.
\end{itemize}
\vspace{-2mm}
\end{minipage}
\hspace{0.01\textwidth}
}

The genus $\infty$ rule is \emph{not} necessary (only sufficient) in general: in \cref{sec:ordertwo}, we discuss the example of $\cG=(\ZZ_4\rtimes\ZZ_4^T)\times\ZZ_2^f$, which has isomorphic SPT classifications despite violating the condition.

The genus $1$ rule, which demands that every valid $\beta$ and $g$ with $x(g)=1$ and $g^2=1$ satisfy $b^*\beta(g)=0$, has a simple meaning in the split case. There, every $\beta$ -- including $\beta=x$ -- is valid, so the rule says that there is no antiunitary symmetry that squares to one. When the group does not split, interpretation of the genus $1$ rule becomes difficult: for example,\footnote{For another example, consider the nonabelian group $\ZZ_4^{Tf}\rtimes\ZZ_2$ and its antiunitary element $g=(1,1)$.} the group $\ZZ_2^T\times\ZZ_4^f$ has antiunitary element $(1,0)$ squaring to $1$ yet satisfies the genus $1$ rule since no $\beta$ that detects this element (in fact, no nonzero $\beta$ at all) is valid.\footnote{That no nontrivial $\beta$ are valid can be seen by writing the $\omega$ for this group and noting that their associated $\nu$ vanish.}

The general result will be proven in \cref{sec:argument} and \cref{sec:genusrules}. First, let us discuss several special cases. One case that we demonstrate in \cref{sec:genusrules} is the following:

\noindent\hspace{0.03\textwidth}\fbox{
\hspace{0.01\textwidth}\begin{minipage}{0.90\textwidth}
\vspace{1mm}
\noindent\emph{Central antiunitary symmetry.}

\vspace{1mm}
When $G_b$ contains a central antiunitary symmetry, the genus $\infty$ rule is necessary and sufficient.
\vspace{1mm}
\end{minipage}
\hspace{0.01\textwidth}
}

Specializing to unitary symmetries, we have the following result, which reveals that the isomorphism for symmetry $\cG=\ZZ_2\times\ZZ_2\times\ZZ_2^f$ considered in \cref{sec:examples} is not an accident but rather an example of the general rule:

\noindent\hspace{0.03\textwidth}\fbox{
\hspace{0.01\textwidth}\begin{minipage}{0.90\textwidth}
\vspace{1mm}
\noindent\emph{Unitary symmetries.}
\vspace{1mm}

When the symmetry $\cG$ is unitary, the groups of bosonic and fermionic SPT phases are isomorphic:
\begin{equation}
    \cB(G_b,\rho,0)\simeq\cF(G_b,\rho,0)~.
\end{equation}

\vspace{-2mm}
\end{minipage}
\hspace{0.01\textwidth}
}

\noindent This result is an immediate corollary of the general result since all of the genus rules are vacuously satisfied when there are no suitable elements $g_i$. It appeared previously in Ref. \cite{Kong}, where it was proven in an abstract setting. In the present paper, we reproduce this result while also stating the isomorphisms explicitly (c.f. \cref{sec:argument}), demystifying its relation to the Jordan-Wigner transformation (which we have seen does \emph{not} typically give an isomorphism even when the groups are isomorphic), studying its relation to more general dualities (c.f. \cref{sec:dualities}), and extending it to include antiunitary symmetries. The unitary result was also stated previously, at least in the split case, in the context of the cobordism group \cite{pontryagindual}. The argument is essentially the one we generalize below: checking that the half cup squares vanish.

Finally, we turn to abelian groups, where conditions directly about the group structure are possible.

\noindent\hspace{0.03\textwidth}\fbox{
\hspace{0.01\textwidth}\begin{minipage}{0.90\textwidth}
\vspace{1mm}
\noindent\emph{Abelian symmetries.}
\vspace{1mm}

When the symmetry $\cG$ is abelian,
\begin{itemize}
    \item The genus $1$ and $2$ rules together are necessary and sufficient.
\end{itemize}

Moreover, when the symmetry $\cG$ is finite abelian and split,
\begin{itemize}
    \item It is necessary and sufficient that $G_b$ is either unitary or has no $\ZZ_2$ factors.
\end{itemize}

\vspace{-2mm}
\end{minipage}
\hspace{0.01\textwidth}
}

\noindent If $\cG$ is unitary, isomorphism of SPT classifications is automatic; if abelian $\cG$ contains an antiunitary symmetry, it is central, and so the genus $\infty$ rule is necessary and sufficient. We observed above that this rule is equivalent to the genus $1$ and $2$ rule together; therefore, rules $1$ and $2$ are necessary and sufficient for abelian symmetries.

It is difficult to translate the genus $n$ rules into statements directly about the group structure -- without reference to $\beta$ -- though we achieve this for finite abelian groups that split. A basis of generators can always be chosen so at most one of them is antiunitary:
\begin{equation}
    G_b=\prod_i\ZZ_{p_i^{n_i}}~,\qquad\text{or}\qquad G_b=\ZZ_{2^n}^T\times\prod_i\ZZ_{p_i^{n_i}}~,
\end{equation}
where the $p_i$ are (not necessarily distinct) primes. The first case, where the symmetry is unitary, is solved. Consider the second case, where one of the generators $t$ is antiunitary. The genus $1$ rules demands there is no valid $\beta$ and antiunitary $g$ with $g^2=1$ and $\beta(g)=1$. The group has an antiunitary $g$ squaring to $1$ if and only if $t^2=1$. Moreover, the $\beta$ that evaluates to $1$ on $t$ and to $0$ on the other generators is valid, since every $\beta$ is valid in the split case. Therefore, the genus $1$ rule is equivalent to $t^2\ne 1$, i.e. $n>1$. Meanwhile, the genus $2$ rule demands that there is no $\beta$ and antiunitary $g_1,g_2$ and such that $g_1^2g_2^2=1$ and $\beta(g_1)\ne\beta(g_2)$. Suppose one of the unitary generators $u$ is of order two, i.e. $p_i=2$, $n_i=1$. Then $g_1=t$ and $g_2=ut^{-1}$ violate the genus $2$ rule with the $\beta$ that evaluates to $1$ on $u$ and to $0$ on the other generators. On the other hand, suppose that none of the unitary generators is of order two, which means that every order two unitary element has a square root. Arbitrary antiunitary elements satisfying $g_1^2g_2^2=1$ are written as $g_1=u_1t^m$ and $g_2=u_2t^{-m}$, which means $(u_1u_2)^2=1$. Then $u_1u_2$ has a square root, so $0=\beta(u_1u_2)=\beta(g_1)+\beta(g_2)$ for any $\beta$. Therefore, the genus $2$ rule is equivalent to the lack of unitary $\ZZ_2$ factors. Together, the genus $1$ and $2$ rules means there are no $\ZZ_2$ factors in $G_b$. The assumption that the symmetry splits is crucial to this conclusion, as demonstrated by the example of $\cG=\ZZ_2^T\times\ZZ_4^f$ mentioned above. Similarly, the example of $\cG=(\ZZ_4\rtimes\ZZ_4^T)\times\ZZ_2^f$ (c.f. \cref{sec:ordertwo}) shows the importance of the assumption that the group is abelian.

We proceed by arguing for the general necessary and sufficient conditions in the following steps:
\begin{enumerate}
    \item[]\label{lemma1} \textbf{Lemma 1.} The groups $\cB$ and $\cF$ are isomorphic if and only if the half cup product is trivial:
    \begin{equation}
        [\tfrac{1}{2}\cup]=0\in H^2(\BB,\AAA)~.
    \end{equation}
    \item[]\label{lemma2} \textbf{Lemma 2.} This happens if and only if, for all $\beta\in\BB$, the half cup square has a square root $R_\beta$ in $\AAA$:
    \begin{equation}
        2R_\beta\sim\tfrac{1}{2}\beta\cup\beta~.
    \end{equation}
    \item[]\label{lemma3} \textbf{Lemma 3.} This happens if the genus $\infty$ rule is satisfied and only if the genus $1$ rule is satisfied.
\end{enumerate}
\hyperref[lemma1]{Lemma 1} was argued in \cref{sec:groupexts}. We turn to \hyperref[lemma2]{Lemma 2} in \cref{sec:argument} and then to \hyperref[lemma3]{Lemma 3} in \cref{sec:genusrules}.

\subsection{Square roots of the diagonal}\label{sec:argument}

Choose a basis $\beta_i$ of generators of $\BB=\ZZ_2^N$. A general valid $\beta$ can then be uniquely decomposed as 
\begin{equation}
    \beta=\sum_ia_i\beta_i~,
\end{equation}
where $a_i$ are $\ZZ_2$-coefficients. In this basis, the half cup squares look like
\begin{equation}
    \tfrac{1}{2}\beta\cup\beta=\tfrac{1}{2}\sum_{ij}a_ia_j\beta_i\cup\beta_j\sim 2\,\tfrac{1}{2}\sum_{i<j}a_ia_j\beta_i\cup\beta_j+\tfrac{1}{2}\sum_ia_ia_i\beta_i\cup\beta_i=\tfrac{1}{2}\sum_ia_i\beta_i\cup\beta_i~.
\end{equation}
For every valid $\beta$, fix a square root $R_{\beta_i}$ of $\tfrac{1}{2}\beta_i\cup\beta_i$ in $\AAA$. A square root of $\tfrac{1}{2}\beta\cup\beta$ is given by
\begin{equation}
    R_\beta=\sum_ia_iR_{\beta_i}
\end{equation}
since
\begin{equation}
    2R_\beta=\sum_ia_i2R_{\beta_i}\sim\tfrac{1}{2}\sum_ia_i\beta_i\cup\beta_i\sim\tfrac{1}{2}\beta\cup\beta~.
\end{equation}
These choices of square roots satisfy
\begin{equation}
    R_{\beta+\beta'}=\sum_i(a_i+a_i'\text{ mod }2)R_{\beta_i}
\end{equation}
and therefore
\begin{equation}
    R_\beta+R_{\beta'}-R_{\beta+\beta'}=\sum_i(a_i+a_i'-(a_i+a_i'\text{ mod }2))R_{\beta_i}=\sum_ia_ia_i'2R_{\beta_i}\sim\tfrac{1}{2}\sum_ia_ia_i'\beta_i\cup\beta_i~.
\end{equation}

Define $\Lambda:\BB\ra\AAA$ as
\begin{equation}\label{lambdaexplicit}
    \Lambda(\beta)=\tfrac{1}{2}\sum_{i<j}a_ia_j\beta_i\cup\beta_j+R_\beta~.
\end{equation}
It has coboundary
\begin{align}\begin{split}\label{dlambda}
    (\delta\Lambda)(\beta,\beta')
    &=\Lambda(\beta)+\Lambda(\beta')-\Lambda(\beta+\beta')\\
    &=\tfrac{1}{2}\sum_{i<j}\left(a_ia_j+a_i'a_j'-(a_i+a_i')(a_j+a_j')\right)\beta_i\cup\beta_j+R_\beta+R_{\beta'}-R_{\beta+\beta'}\\
    &\sim\tfrac{1}{2}\sum_{i<j}\left(a_ia_j'+a_i'a_j\right)\beta_i\cup\beta_j+\tfrac{1}{2}\sum_ia_ia_i'\beta_i\cup\beta_i\\
    &=\tfrac{1}{2}\sum_{i,j}a_ia_j'\beta_i\cup\beta_j\\
    &=\Omega_\cF(\beta,\beta')~.
\end{split}\end{align}
This means that, given the existence of square roots $R_\beta$, the extension class $[\Omega_\cF]=[\tfrac{1}{2}\cup]$ is trivial. Conversely, suppose $\Omega_\cF=\delta\Lambda$ for some $\Lambda$, not necessarily of the form \eqref{lambdaexplicit}. The diagonal terms are
\begin{equation}
    [\tfrac{1}{2}\beta\cup\beta]=\Omega_\cF(\beta,\beta)=(\delta\Lambda)(\beta,\beta)=2\Lambda(\beta)
\end{equation}
since $\Lambda(0)=\Omega_F(\beta,0)=0$, so $\Lambda(\beta)$ is a square root of $\tfrac{1}{2}\beta\cup\beta$ in $\AAA$. This proves \hyperref[lemma2]{Lemma 2}.

\subsection{Derivation of the genus rules}\label{sec:genusrules}

Next is \hyperref[lemma3]{Lemma 3}, which states a property of the symmetry group that guarantees that its valid half cup squares have square roots (and therefore, by the previous lemmas, that the SPT classification groups are isomorphic). A special case occurs when these half cup squares \emph{vanish} in $\AAA$ -- if they vanish, they have zero as a square root. We argue in the following that the \hyperref[genusrule]{genus $\infty$ rule} is equivalent to them vanishing. While this condition is sufficient for the existence of square roots (namely, zero), it isn't necessary, as a nonvanishing valid half cup square may still have a square root (see the example of $G_b=\ZZ_4\rtimes\ZZ_4^T$ in \cref{sec:ordertwo}). The genus $1$ condition, on the other hand, must be satisfied for the valid half cup squares to have square roots. 

In turns out that when there is an antiunitary symmetry in the center of $G_b$, all classes in $\AAA$ are of order two, which means they only have square roots if they vanish. In this case, the \hyperref[genusrule]{genus $\infty$ rule} is necessary as well as sufficient. We prove the claim about classes of order two in \cref{sec:ordertwo}. 

We proceed by studying when the valid half cup squares vanish in $\AAA$. Since $[\tfrac{1}{2}\rho]$ generates the kernel of $b^*:H^2(G_b,U(1)_x)\ra H^2(\cG,U(1)_x)$, the vanishing of a class in $\AAA=H^2(G_b,U(1)_x)/\langle\tfrac{1}{2}\rho\rangle$ is equivalent to the vanishing of its pullback by $b$ in $H^2(\cG,U(1)_x)$. Therefore, we wish to know when
\begin{equation}\label{cupsquareomega}
    \omega=\tfrac{1}{2}b^*\beta\cup b^*\beta
\end{equation}
is trivial as a $\cG$ cocycle, for every valid $\beta$. To study this, we employ the bosonic partition functions \eqref{bosonictheory} associated with this cocycle. Partition functions are gauge invariant quantities, which makes them useful for studying the cohomology class of a cocycle. Moreover, partition functions of unitary, invertible theories like ours completely determine\footnote{The complete list of partition functions is redundant. See e.g. relations between tori, Klein bottles, and $\RR P^2$ in Ref. \cite{ShiozakiRyu}.} their topological invariants in the sense that two invariants $[\omega]$ and $[\omega']$ are equal if and only if their partition functions $\cZ_\omega^b$ and $\cZ_{\omega'}^b$ agree on every space $(X,\cA)$ \cite{yonekura,freedmoore}. Therefore, asking when an invariant is trivial amounts to asking when its partition functions are all trivial.

The bosonic partition function associated with a cocycle $\omega$ is stated abstractly in Eq. \eqref{bosonictheory}. Now we evaluate this expression explicitly for each input $(X,\cA)$. The surface $X$ is either orientable -- in which case it is the connected sum of $n$ tori -- or nonorientable -- in which case it is the connected sum of $n$ real projective planes; in either case, $n$ is referred to as the ``genus.'' The genus $n$ orientable surface has fundamental group
\begin{equation}\label{orientablefund}
    \pi_1((T^2)^{\#n})=\{\,a_i,b_i~,\,i=1,\ldots,n\,:\,\prod_{i}^na_ib_ia_i^{-1}b_i^{-1}=1\,\}
\end{equation}
generated by \emph{orientation-preserving} loops $a_i, b_i$, whereas the genus $n$ nonorientable surface has
\begin{equation}\label{nonorientablefund}
    \pi_1((\RR P^2)^{\#n})=\{\,c_i~,\,i=1,\ldots,n\,:\,\prod_{i}^nc_i^2=1\,\}
\end{equation}
generated by \emph{orientation-reversing} loops $c_i$ \cite{hatcher}. The gauge field $\cA$ is determined by the map $\pi_1(X)\ra\cG$ induced on fundamental groups; that is, by specifying group elements (``holonomies'') $\{g_i,h_i\}$ or $\{g_i\}$ that satisfy the relation \eqref{orientablefund} or \eqref{nonorientablefund} and the constraint $\cA^*x=w_1$, which means $x(g_i)=x(h_i)=0$ or $x(g_i)=1$.

Next we provide two independent arguments that the partition functions associated with the class \eqref{cupsquareomega} automatically vanish on orientable surfaces
\begin{equation}\label{orientableonbub}
    \cZ^b_{\tfrac{1}{2}b^*\beta\cup b^*\beta}[(T^2)^{\#n},\{g_i,h_i\}]=1
\end{equation}
and evaluate to
\begin{equation}\label{nonorientableonbub}
    \cZ^b_{\tfrac{1}{2}b^*\beta\cup b^*\beta}[(\RR P^2)^{\#n},\{g_i\}]=(-1)^{\sum_ib^*\beta(g_i)}~.
\end{equation}
on the nonorientable surface with holonomies $\{g_i\}$. The condition that the nonorientable genus $n$ partition function vanishes for all valid $\beta$ is precisely the \hyperref[genusrule]{genus $n$ rule} and the vanishing for all $n$ is the genus $\infty$ rule. Therefore the genus $\infty$ rule is a sufficient condition for the existence of square roots $R_\beta$ to the valid half cup squares and thus to the isomorphism of bosonic and fermionic SPT classifications.

Our first argument uses the special form of the class \eqref{cupsquareomega}, which lets us write
\begin{equation}
    \cZ^b_{\tfrac{1}{2}b^*\beta\cup b^*\beta}[X,\cA]=\exp(2\pi i\int_X \tfrac{1}{2}A_b^*\beta\cup A_b^*\beta)=(-1)^{\langle PD(A_b^*\beta),PD(A_b^*\beta)\rangle}~,
\end{equation}
where $PD(A_b^*\beta)$ denotes the cycle Poincar\'{e} dual to $A_b^*\beta\in H^1(X,\ZZ_2)$ and $\langle\cdot,\cdot\rangle$ is the intersection form on cycles. On an orientable surface, no cycle has even self-intersection \cite{hatcher}, so the partition function is trivial. On the other hand, the orientation-reversing loops $c_i$ in the presentation \eqref{nonorientablefund} form a basis of cycles of the nonorientable surface and have intersection $\langle c_i,c_j\rangle=\delta_{ij}$ \cite{hatcher}; therefore, the self-intersection of any cycle is the number of basis elements $c_i$ along which the cycle has a component. The Poincar\'{e} dual cycle $PD(A_b^*\beta)$ is, by definition, the cycle that intersects precisely the cycles $c$ on which $A_b^*\beta(c)=1$; since the basis $c_i$ is orthogonal with respect to the intersection form, the Poincar\'{e} dual therefore has components precisely where it has $A_b^*\beta(c_i)=1$, i.e. in the basis cycles labeled by holonomies $g_i\in\cG$ such that $b^*\beta(g_i)=1$. We conclude that the self-intersection of $PD(A_b^*\beta)$ is a sum of $b^*\beta(g_i)$, which gives the claimed result \eqref{nonorientableonbub}.

Our second argument uses the general expression for the partition function associated to $\omega$ evaluated on a space $(X,\cA)$. In \cref{sec:statesum}, we derive these expressions from the state sum construction and compare them to partial results in the literature. We find that the orientable partition function is
\begin{align}\begin{split}\label{orientableevaluated}
    \cZ^b_\omega[(T^2)^{\#n},\{g_i,h_i\}]
    &=\exp\left(2\pi i\sum_{i=1}^n\omega(x_i,g_i)+\omega(x_ig_i,h_i)+\omega(x_ig_ih_i,g_i^{-1})\right.\\
    &\left.\qquad\qquad\qquad\qquad\qquad\phantom{\sum_{i=1}^n}+\omega(x_ig_ih_ig_i^{-1},h_i^{-1})-\omega(g_i^{-1},g_i)-\omega(h_i^{-1},h_i)\right)~,
\end{split}\end{align}
for $\{g_i,h_i\}$ satisfying $x(g_i)=x(h_i)=0$ and $\prod_ig_ih_ig_i^{-1}h_i^{-1}=1$, and where $x_i=\prod_{j<i}g_jh_jg_j^{-1}h_j^{-1}$. And for $\{g_i\}$ satisfying $x(g_i)=1$ and $\prod_i g_i^2=1$, with $y_i=\prod_{j<i}g_j^2$, the nonorientable partition function is
\begin{equation}\label{nonorientableevaluated}
    \cZ^b_\omega[(\RR P^2)^{\#n},\{g_i\}]=\exp(2\pi i\sum_{i=1}^n \omega(g_i,g_i)+\omega(y_i,g_i^2))~.
\end{equation}
Gauge invariance of these expressions can be verified by plugging in $\omega=\delta\lambda$ to obtain
\begin{equation}
    \cZ^b_{\delta\lambda}[(T^2)^{\#n},\{g_i,h_i\}]=\exp(2\pi i\sum_{i=1}^n\lambda(x_i)+\lambda(x_{i+1}))=1~,
\end{equation}
\begin{equation}
    \cZ^b_{\delta\lambda}[(\RR P^2)^{\#n},\{g_i\}]=\exp(2\pi i\sum_{i=1}^n \lambda(y_i)+\lambda(y_{i+1}))=1~,
\end{equation}
using the facts that $x_1=1$, $x_ng_nh_ng_n^{-1}h_n^{-1}=1$, $y_1=1$, and $y_ng_n^2=1$. Plugging the class \eqref{cupsquareomega} into the general evaluations \eqref{orientableevaluated} and \eqref{nonorientableevaluated} and noting that $b^*\beta(x_i)=b^*\beta(y_i)=0$, we see that the orientable expression becomes trivial while the nonorientable expression matches the claimed result \eqref{nonorientableonbub}.

To see the partial converse, note that the partition function on $\RR P^2$ for an arbitrary class $[\omega]$ is of order two since
\begin{equation}
    \cZ^b_\omega[\RR P^2_g]^2=\exp(2\pi i\,2\omega(g,g))=\exp(2\pi i\,(\delta\omega)(g,g,g))=1~,
\end{equation}
due to the fundamental group relation $g^2=1$. For this reason, a class with nontrivial $\RR P^2$ partition function cannot have a square root. We conclude that a square root $R_\beta$ exists only if the partition function \eqref{nonorientableonbub} vanishes at $n=1$; that is, if the \hyperref[genusrule]{genus $1$ rule} is satisfied.

\section{Dualities of SPT phases}\label{sec:dualities}

\subsection{Topological holography}

Duality is a tool for relating two systems via a nontrivial transformation such that there is a correspondence between the key physical observables on either side. Such transformations allow for solving problems about one system in terms of its dual. For example, the Jordan-Wigner transformation \cite{JordanWigner}, which maps nonlocal strings of bosonic spin operators to local fermion operators, has proven useful for solving spin chains such as the Ising and XY models by translating them into fermionic Hamiltonians that can be diagonalized \cite{LSM,derzhko2001jordanwigner}. The Jordan-Wigner transformation has also been used to study fermionic SPT phases by mapping them into bosonic SPT phases \cite{CGW_complete,FidkowskiKitaev}. Other examples are Kennedy-Tasaki duality \cite{KT} and its generalizations \cite{Else_2013,li2023noninvertible}, which interpret the nontriviality of SPT phases as ``hidden'' symmetry-breaking by mapping SPT phases to spontaneously symmetry-broken phases. Dualities can also be used to locate critical points as points where a duality between two phases becomes a self-duality, as was done by Kramers and Wannier \cite{KW}.

While the specific duality transformations mentioned above have appeared in separate contexts, they can be understood as instances of a general notion of duality \cite{Freed_2022, aksoy2023liebschultzmattis,li2023noninvertible}. Recent years have shown progress in understanding duality from within a unified framework known as topological holography, where symmetric gapped phases are realized as gapped boundary conditions of higher dimensional symmetry topological field theories (TFT). This idea has proven especially fruitful in one dimension, where systems with finite symmetry\footnote{The approach applies not only to the grouplike symmetries considered here but also to their noninvertible generalizations.} and their dualities have been classified in the language of two dimensional symmetry TFT \cite{Gaiotto_2021,Freed_2022,Lichtman_2021,moradi2022topological,freed2023topological, bhardwaj2023generalized}. The symmetries of one dimensional systems form a fusion category $\cA$ (for the nonanomalous grouplike symmetries considered in the present paper, $\cA=\text{Vec}_\cG$) \cite{Bhardwaj_2018, Chang_2019}, and the symmetry TFT is the two dimensional theory characterized by the Drinfeld center $Z(\cA).$ One dimensional $\cA$-symmetric phases (including both symmetry-protected and symmetry-breaking phases) appear as gapped boundary conditions for $Z(\cA)$, which are mathematically described by Lagrangian algebra objects of $Z(\cA)$, consisting of the anyons that condense on the boundary \cite{thorngren2019fusion, Kong_2014}. Dualities of one dimensional systems (including operations like gauging a subgroup and stacking with SPT phases) appear in this framework as braided autoequivalences of the modular tensor category $Z(\cA)$ -- that is, as permutations of the anyons that preserve their braiding and fusion rules -- which act on the Lagrangian algebras representing the phases \cite{Barkeshli_2019,Jia_2023}. These autoequivalences may be physically interpreted as invertible surface operators (also called gapped domain walls or gapped interfaces) of the two dimensional symmetry TFT \cite{kapustin2010surface, Gaiotto_2021}.

It is expected that the dualities formulated in the symmetry TFT language are related to duality transformations of lattice Hamiltonians. For example, the electric-magnetic duality of the toric code TFT, which exchanges the $e$ and $m$ anyons, is known appear as gauging and Kramers-Wannier duality on the one dimensional lattice. Dualities of systems with nonanomalous abelian symmetry groups are extensively studied in Ref. \cite{moradi2022topological}, where dual pairs of lattice Hamiltonians are constructed for a given autoequivalence. See Refs. \cite{lootens2022dualities,Lootens_2023} for a construction of dual pairs of lattice models for more general symmetries. The Jordan-Wigner transformation has also been discussed as a duality in the framework of symmetry TFT, though a general formalism for boson-to-fermion dualities has yet to be fully developed \cite{Gaiotto_2021,lootens2022dualities}. 

In \cref{sec:isomorphic} it was found that the stacking rules of bosonic and fermionic SPT phases are often isomorphic; this happens when the symmetry is unitary as well as when there are antiunitary symmetries as long as certain conditions are satisfied (c.f. \cref{sec:necandsuf}). In these cases, an explicit isomorphism $\cX$ \eqref{isommap} between the groups of bosonic and fermionic SPT phases was constructed. In this section we ask whether the transformation $\cX$ -- like the Jordan-Wigner transformation -- can be thought of as a duality, in the sense of being implemented by an autoequivalence of the symmetry TFT. If true, this would mean that boson-to-fermion dualities are capable of preserving stacking rules, which are properties of entire collections of SPT phases, similarly to how they preserve key features of individual phases when mapping from one system to another. We show however that $\cX$ is \emph{not} a duality, except in the special case where the Jordan-Wigner transformation is an isomorphism of stacking rules ($\delta\Lambda=\Omega_\cF=0$). We will focus on the case where $\cG$ is finite abelian and unitary, but we expect the spirit of the arguments to apply more broadly. In order to make use of the symmetry TFT framework, which is typically formulated only for bosonic systems, we consider the transformation
\begin{equation}
    \cJ\cW^{-1}\cX:\cB\ra\cB
\end{equation}
between bosonic SPT phases. Since $\cJ\cW$ may be regarded as a duality, asking whether $\cX$ is a duality between bosonic and fermionic systems is equivalent to asking whether $\cJ\cW^{-1}\cX$ is a duality. We emphasize that, while $\cX$ is by construction an isomorphism $\cB\ra\cF$, the composite transformation $\cJ\cW^{-1}\cX$ is typically not an automorphism of the group of bosonic SPT phases since $\cJ\cW$ is not typically an isomorphism.

\subsection{Actions of dualities on SPT invariants}
\label{sec:dualities_Lag_SPT}

Anyons of abelian $Z({\rm Vec}_\cG)$ take the form $(a,\chi)$, where $a \in \cG$ represents a magnetic flux and $\chi \in \Rep(\cG) \simeq \cG$ an electric charge, with the topological spin of $(a,\chi)$ given by the evaluation $\theta_{(a, \chi)} = \chi(a)$. Anyons of the form $(1,\chi)$ are pure electric charges and form the subcategory $\Rep(\cG)\subset Z({\rm Vec}_\cG)$.

Gapped phases with $\cG$ symmetry correspond to Lagrangian algebras of $Z({\rm Vec}_\cG)$. These are classified by pairs $(H,\omega)$, where $H$ is a subgroup of $\cG$ representing the unbroken symmetry and $\omega\in H^2(H,U(1))$ encodes the SPT order. The Lagrangian algebra labeled by $(H,\omega)$ is, as an object,\footnote{In general the Lagrangian algebra structure on such an object may not be unique, so the object alone may not determine the class $\omega$; for abelian groups, however, the underlying object fixes the class completely.} the direct sum of anyons
\begin{equation}
    \cL_{H,\omega}=\bigoplus_{\substack{a\in H~,\,\,\chi\in\text{Rep}(\cG) \\ \chi|_H=(i_a\omega)|_H}}(a,\chi)~,
\end{equation}
where $|_H$ denotes restriction to $H$ and $i_a\omega$ denotes the representation given by the slant product
\begin{equation}
    (i_a\omega)(g)=\exp(2\pi i(\omega(a,g)-\omega(g,a)))
\end{equation}
of $\omega$ with respect to $a\in H$. SPT phases correspond to the case $H=\cG$ and are classified by $\omega\in H^2(\cG, U(1))$. The corresponding Lagrangian algebras
\begin{equation}\label{eq:SPT-slant}
    \cL_{\cG,\omega}=\bigoplus_{a\in \cG}(a,i_a\omega)
\end{equation}
involve no pure electric charges $(1,\chi)$ since $i_1\omega=1$ on $\cG$; they are thus referred to as ``magnetic'' \cite{zhang2023anomalies}. The Lagrangian algebra of the trivial SPT phase involves only pure magnetic fluxes since $i_a0=1$:
\begin{equation}
    \cL_{\cG,0}=\bigoplus_{a\in \cG}(a,1)~.
\end{equation}
A physical picture of the correspondence between SPT invariants and Lagrangian algebras may be given in terms of order parameters \cite{Lichtman_2021, bhardwaj2023categorical, bhardwaj2023gapped}.

Let us illustrate this story with a simple example: $\cG = \ZZ_2^x \times \ZZ_2^y$. Let $x$ and $y$ denote the generators of the $\ZZ_2$ factors. In this basis, the anyons generating $Z({\rm Vec}_G)$ are
\begin{equation}\label{anyongens}
    m_x=((1,0),(0,0))~,\qquad m_y=((0,1),(0,0))~,\qquad e_x=((0,0),(1,0))~,\qquad e_y=((0,0),(0,1))~.
\end{equation}
The trivial SPT phase corresponds to $\omega= 0$ and involves only pure magnetic fluxes:
\begin{equation}
    \cL_{\cG, 0}=1 \oplus m_x \oplus m_y \oplus m_x m_y.
    \label{eq:lag_triv}
\end{equation} 
The nontrivial SPT phase corresponds to $\omega = \omega_{xy}$, the nontrivial class in $H^2(\ZZ_2 \times \ZZ_2, U(1)).$ This class has
\begin{equation}
    i_x\omega_{xy} = e_y~,\qquad i_y\omega_{xy} = e_x~.
\end{equation}
Therefore the corresponding Lagrangian algebra is
\begin{equation}
    \cL_{\cG, \omega_{xy}} = 1 \oplus m_x e_y \oplus m_y e_x \oplus m_x m_y e_xe_y,
    \label{eq:lag_SPT}
\end{equation}
reproducing the well-known expression for the $\ZZ_2 \times \ZZ_2$ SPT-phase, for example in Ref. \cite{moradi2022topological}.

The correspondence between SPT invariants $\omega$ and Lagrangian algebras allows us to compute the action of braided autoequivalences $\cD\in\Aut(Z({\rm Vec}_\cG))$ on SPT phases and address the question of whether the map
\begin{equation}\label{jwinvX}
    \cJ\cW^{-1}\cX:[\alpha,\beta]\mapsto[\alpha-\Lambda(\beta),\beta]~,
\end{equation}
coming from Eq. \eqref{isommap}, can be implemented by such an autoequivalence.

Before getting to the general result, consider the special case of dualities which map pure electric charges to pure electric charges; that is, which stabilize $\Rep(\cG)$. For symmetries such as $\cG=\ZZ_2^n$, where every non-charge boson appears in some SPT phase, every duality that maps SPT phases to SPT phases (and so stabilizes the set of non-charge bosons) stabilizes $\Rep(\cG)$. These dualities are known to act on SPT phases by stacking with an SPT phase $\zeta$, then acting by the map $a^*$ induced by an automorphism $a\in{\rm Out}(\cG)$:\footnote{For the abelian groups considered here, every automorphism is of course outer. More generally, the fact that the automorphism $a$ is outer is reflected in Eq. \eqref{conjgauge}, which says that inner automorphisms act trivially on the SPT invariants.}
\begin{equation}\label{eq:repGstab-action}
    \omega\mapsto a^*\omega+\zeta~.
\end{equation}
This fact appears in Ref. \cite{nikshych2013categorical} (Cor. 6.9, Prop 7.11),\footnote{Ref. \cite{nikshych2013categorical} considers the action on Lagrangian subcategories, but Lagrangian subcategories and Lagrangian algebras coincide on the level of anyons when $\cG$ is abelian -- Lagrangian subcategories are fusion categories generated by some subset of anyons of $Z({\rm Vec}_\cG)$, while Lagrangian algebras correspond to direct sums of the same anyons \cite{Naidu_2008}.} and we give a self-contained proof of it in \cref{sec:dualities_app}. Now let us ask when $\cJ\cW^{-1}\cX$ is of this form. It fixes the trivial phase, so the stacked SPT $\zeta$ must be trivial. The induced action $a^*$ on the SPT classification is an automorphism, $a^*(\omega+\omega')=a^*\omega+a^*\omega'$, so $\cJ\cW^{-1}\cX$ is only of this form when it is an automorphism. Since $\cX$ is an isomorphism, $\cJ\cW$ must be an isomorphism as well, which it typically is not. This means there is no duality transformation implementing $\cX$ in general.

It turns out that the full theory of dualities that map SPT phases to SPT phases is more complicated, as not all of them stabilize $\Rep(\cG)$ or even act on SPT phases like those that stabilize $\Rep(\cG)$; in fact, there are $\cL_{\cG,0}$-fixing, SPT-stabilizing dualities whose actions on SPT phases are not automorphisms. Nevertheless, the conclusion of the previous paragraph holds. In \cref{sec:dualities_app}, we demonstrate through examples how dualities can induce exotic actions on SPT phases and we prove the following theorem:

\noindent\hspace{0.03\textwidth}\fbox{
\hspace{0.01\textwidth}\begin{minipage}{0.90\textwidth}
\vspace{1mm}
\textbf{Theorem:} If the transformation $\cJ\cW^{-1}\cX$ of SPT phases is implemented by a duality, then $\cJ\cW$ is an isomorphism of the bosonic and fermionic stacking groups, i.e. $\Omega_\cF=\delta\Lambda=0$.
\vspace{1mm}
\end{minipage}
\hspace{0.01\textwidth}
}

\subsection{Stacking of magnetic Lagrangian algebras} \label{sec:dualities_Lag_Stack}

In \cref{sec:dualities_Lag_SPT} we studied the interplay of stacking and duality by expressing both in terms of the SPT invariant $\omega$. In this picture, stacking is straightforward, while the action of autoequivalences is not. Here we consider an alternative approach, where the action of autoequivalences is immediate but stacking needs to be carefully defined. The idea is to express everything in terms of anyons and Lagrangian algebras. In this section we define the operation of stacking Lagrangian algebras and see how it plays out in the case of $\cG = \ZZ_2 \times \ZZ_2$.

Given two Lagrangian algebras $\cL_1, \cL_2$ of $Z({\rm Vec}_\cG)$, thought of as gapped boundaries of the bulk symmetry TFT, we can stack the whole bulk-boundary systems to obtain a Lagrangian algebra $\cL_1 \boxtimes \cL_2$ of the bulk $Z({\rm Vec}_\cG) \boxtimes Z({\rm Vec}_\cG)$, where $\boxtimes$ denotes the Deligne product. The bulk is now a symmetry TFT for $\cG \times \cG$ symmetry, which we must break into the diagonal subgroup $\cG$ by identifying the $\Rep(\cG)$ subcategories of the two factors. This is achieved by condensing the algebra object
\begin{equation}\label{condensable}
    \cE = \bigoplus_{\alpha \in \Rep(\cG)} \alpha \boxtimes \alpha
\end{equation}
in $Z({\rm Vec}_\cG) \boxtimes Z({\rm Vec}_\cG)$ \cite{Lan_2017}. The result of this condensation, denoted $\left( Z({\rm Vec}_\cG) \boxtimes Z({\rm Vec}_\cG) \right)_{\cE}$, is equivalent to and identified with $Z({\rm Vec}_\cG)$ \cite{Lan_2016, Lan_2017}. The stacked Lagrangian algebra\footnote{This procedure defines the stacking of two general Lagrangian algebras, not necessarily magnetic. Non-magnetic Lagrangian algebras correspond to (partially) symmetry-breaking phases, and the stacking of these phases is indeed well-defined; they simply will not form a group under stacking, but a commutative monoid with some noninvertible elements.} is defined as the result of forming the product of the Lagrangian algebras, then condensing $\cE$, and finally identifying the result with an object of $Z({\rm Vec}_\cG)$:
\begin{equation}
    \cL_1 \cdot \cL_2 := \left( \cL_1 \boxtimes \cL_2 \right)_{\cE}~.
\end{equation}

Let us illustrate this procedure concretely by returning to the example of $\cG = \ZZ_2^x \times \ZZ_2^y$. The $16$ anyons in $Z({\rm Vec}_\cG)$ are generated by $e_x, e_y, m_x, m_y$ \eqref{anyongens}, and the $16^2$ anyons of $Z({\rm Vec}_\cG) \boxtimes Z({\rm Vec}_\cG)$ are generated by the Deligne products of $e_x, e_y, m_x, m_y$. The condensable algebra \eqref{condensable} is given by
\begin{equation}
    \cE = (1 \boxtimes 1) \oplus (e_x \boxtimes e_x) \oplus (e_y \boxtimes e_y) \oplus (e_xe_y \boxtimes e_x e_y)~. 
\end{equation}
The identification $\left( Z({\rm Vec}_\cG) \boxtimes Z({\rm Vec}_\cG)\right)_{\cE} \simeq Z({\rm Vec}_\cG)$ can be made as follows. After condensing $\cE$, we have relations $e_i \boxtimes 1 \sim 1 \boxtimes e_i \sim e_i e_j \boxtimes e_j$, and so forth. Let $[e_i]$ be the class to which $e_i \boxtimes 1$ belongs. This way, we obtain four classes with trivial magnetic flux, $[1], [e_x], [e_y], [e_x e_y]$. As for the anyons involving nontrivial magnetic fluxes, $m_i \boxtimes m_j$ has nontrivial braiding with at least one of the condensed anyons if $i \neq j$, and are confined. Hence deconfined magnetic anyons take the diagonal form $m_i \boxtimes m_i$. These are then identified via condensation as $m_i \boxtimes m_i \sim m_i e_x \boxtimes m_i e_x$, etc. Let $[m_i]$ denote the equivalence class to which $m_i\boxtimes m_i$ belongs. Equivalence classes of deconfined anyons of $Z({\rm Vec}_\cG) \boxtimes Z({\rm Vec}_\cG)$ are generated by the equivalence classes $[e_i]$ and $[m_i]$. We then identify $[e_i]$ and $[m_i]$ with the anyons $e_i$ and $m_i$, respectively, of  $Z({\rm Vec}_\cG)$.

Now consider the stacking of SPT phases in this example. The Lagrangian algebras corresponding to the trivial and nontrivial $\cG=\ZZ_2\times\ZZ_2$ SPT phases were stated in Eqs. \eqref{eq:lag_triv} and \eqref{eq:lag_SPT}. We wish to compute the result of stacking the algebra $\cL_{\cG, \omega_{xy}}$ with itself. Consider the $16$ anyons of $\cL_{\cG, \omega_{xy}} \boxtimes \cL_{\cG, \omega_{xy}}$. The anyon $m_x e_y \boxtimes m_x e_y$ belongs to the class $[m_x]$, and thus maps to the anyon $m_x$ of $Z(\text{Vec}_\cG)$ after condensation. Similarly, $m_y e_x \boxtimes m_y e_x$ maps to $m_y$. Anyons such as $m_x e_y \boxtimes m_y e_x$ are confined. Thus we see that 
\begin{equation}
    \cL_{\cG, \omega_{xy}}\cdot \cL_{\cG, \omega_{xy}} =\left( \cL_{\cG, \omega_{xy}}\boxtimes \cL_{\cG, \omega_{xy}}\right)_{\cE} = 1 \oplus m_x \oplus m_y \oplus m_x m_y = \cL_{\cG, 0}~,
\end{equation}
the Lagrangian algebra corresponding to the trivial phase. Thus, we see that the stacking rule of $\ZZ_2\times\ZZ_2$ SPT phases is recovered from the Lagrangian algebra perspective.

Now that we have made sense of stacking in the language of anyons, it is simple to study dualities. For example, consider the autoequivalence
\begin{equation}
    \cD:\quad m_x \longleftrightarrow m_y~,\quad e_x\longleftrightarrow e_y~.
\end{equation}
This autoequivalence is induced from an automorphism of $\cG$ (as pure magnetic fluxes correspond to group elements) and leaves each of $\cL_{\cG, 0}$ and $\cL_{\cG,\omega_{xy}}$ invariant, as can be seen from the expressions \eqref{eq:lag_triv} and \eqref{eq:lag_SPT}. Or consider the autoequivalence
\begin{equation}
    \cD:\quad m_i \longleftrightarrow m_i e_j\,,\quad j \neq i~,
\end{equation}
which exchanges $\cL_{\cG, 0}$ and  $\cL_{\cG, \omega_{xy}}$ and thus corresponds to stacking with the nontrivial SPT phase. To study how an autoequivalence $\cD\in\Aut(Z({\rm Vec}_\cG))$ interacts with the stacking of Lagrangian algebras, one should compare two objects: first, the result of stacking then dualizing
\begin{equation}
    \cD(\cL_1 \cdot \cL_2) = \cD \left( (\cL_1 \boxtimes \cL_2 )_\mathcal{E} \right)~,
\end{equation}
and second, the result of dualizing then stacking
\begin{equation}
    \cD(\cL_1) \cdot  \cD(\cL_2) = \left( \cD(\cL_1) \boxtimes  \cD(\cL_2)  \right)_{\cD(\cE)}~.
\end{equation}
Note that the condensable algebra $\cE$ becomes the dualized object $\cD(\cE)$ in this second procedure.

\subsection{An example, revisited}\label{dualities:example}

Finally we return to the example of SPT phases with symmetry group $\cG=\ZZ_2\times\ZZ_2\times\ZZ_2^f$, first examined in \cref{sec:examples}. Recall that here the stacking rules of bosonic and fermionic SPT phases are isomorphic despite the Jordan-Wigner transformation not being an isomorphism. It was natural before to wonder whether the isomorphism $\cX$ could, like $\cJ\cW$, be implemented by a duality transformation. We have since seen in this section that the answer is no. Here we recover this result by working explicitly with the action of dualities on the $64$ anyons and on the $8$ magnetic Lagrangian algebras corresponding to the $8$ SPT phases.

\begin{table}[!b]
    \centering
   $\begin{tabu}{c|c|lll}
\alpha~,\,\beta & \omega & \multicolumn{3}{c}{\text{generating set of anyons}}  \\
\hline 
 \alpha_0~,\,\beta_{00} & 0 & m_x &  m_y &   m_p  \\ 
 \alpha_0~,\,\beta_{10} &   \omega_{xp} &  m_x e_p & m_y & m_p e_x 
  \\
  \alpha_0~,\,\beta_{01} &   \omega_{yp}  &   m_x & m_y e_p & m_p e_y  \\
  \alpha_0~,\,\beta_{11} &     \omega_{xp} + \omega_{yp} &  m_x e_p & m_y e_p & m_p e_x e_y  \\
\alpha_1~,\,\beta_{00} & \omega_{xy} & m_x e_y & m_y e_x & m_p  \\  
\alpha_1~,\,\beta_{10} &  \omega_{xy} + \omega_{xp} &  m_x e_y e_p & m_y e_x & m_p e_x  \\
 \alpha_1~,\,\beta_{01} &      \omega_{xy}  +\omega_{yp}  &  m_x e_y & m_y e_x e_p & m_p e_y \\
\alpha_1~,\,\beta_{11} &  \omega_{xy} + \omega_{xp}  + \omega_{yp} &  m_x e_y e_p & m_y e_x e_p & m_p e_x e_y 
\end{tabu}$
    \caption{The $8$ SPT phases with symmetry $\cG = \ZZ_2 \times \ZZ_2 \times \ZZ_2^f$.}
    \label{tab:LagAlg_example}
\end{table}

As discussed in \cref{sec:examples}, the $8$ SPT phases are given by a choice of $\alpha$ between $\alpha_0$ and $\alpha_1$ and a choice of $\beta$ among $\beta_{00}$, $\beta_{10}$, $\beta_{01}$, and $\beta_{11}$. The map $\cJ\cW^{-1}\cX$ exchanges the two phases
\begin{equation}\label{exampleduality}
    \cJ\cW^{-1}\cX:[\alpha_0,\beta_{11}]\longleftrightarrow[\alpha_1,\beta_{11}]~,
\end{equation}
while leaving the other six phases the same. We will see that no duality can implement this transformation.

The relevant symmetry TFT is the rank $64$ abelian MTC $Z(\rm{Vec}_{\ZZ_2 \times \ZZ_2 \times \ZZ_2}) = {\rm TC} \boxtimes {\rm TC} \boxtimes {\rm TC}$, where ${\rm TC}$ denotes the Toric Code MTC. This MTC has $\ZZ_2^6$ fusion rules generated by the anyons $e_x, e_y, e_p, m_x, m_y, m_p$ and has $8$ magnetic Lagrangian algebras. The correspondence between these algebras and the $8$ SPT phases is computed following the procedure discussed in \cref{sec:dualities_Lag_SPT}. The results appear in \cref{tab:LagAlg_example}, where each Lagrangian algebra is denoted by its set of generating anyons. These algebras also appear in Ref. \cite{moradi2022topological} (c.f. Table 6).

Suppose the transformation $\cJ\cW^{-1}\cX$ \eqref{exampleduality} were implemented by a duality. The invariance of the trivial phase $(\alpha_0,\beta_{00})$ under this mapping means that the pure magnetic fluxes must map to other pure magnetic fluxes. Then, by looking at the table, we see that the invariance of the phase $(\alpha_0,\beta_{10})$ means the anyon $m_y$ must be invariant since it is the only pure magnetic flux for this phase. Likewise, the invariance of the phase $(\alpha_0,\beta_{01})$ implies the invariance of $m_x$. Therefore the anyon $m_xm_y$ is invariant. However the phase $(\alpha_0,\beta_{11})$ has the anyon $m_xm_y=(m_xe_p)(m_ye_p)$ while the phase $(\alpha_1,\beta_{11})$ does not, so they cannot be mapped into each other by the transformation, a contradiction. We conclude that the transformation $\cJ\cW^{-1}\cX$ -- and thus also the stacking rule isomorphism $\cX:\cB\ra\cF$ -- is not implemented by a duality.

\section{Discussion}

We have investigated the conditions under which one dimensional bosonic and fermionic SPT phases have isomorphic stacking rules, finding that this occurs when the symmetry group is purely unitary and extending this result to antiunitary symmetries, where it happens in certain circumstances. By thinking of the bosonic and fermionic SPT classifications as group extensions that are sometimes equivalent, we have found several necessary and sufficient conditions for an isomorphism to hold; these results are gathered in \cref{sec:necandsuf}. A natural question is whether an isomorphism between the groups of bosonic and fermionic SPT phases might exist in higher dimensions. There, the situation is complicated by the fact that gauging the fermion parity symmetry of a fermionic system gives rise to an anomalous higher form $\ZZ_2$ symmetry of the resulting bosonic theory \cite{GK,Bhardwaj_2017}, rather than to the nonanomalous ordinary $\ZZ_2$ symmetry of one dimension, so we do not expect the invertibility of phases to be preserved by this operation. Indeed, it is known that in two dimensions the standard bosonization procedure maps an invertible fermionic phase to a bosonic topologically-ordered phase, and this fact has been exploited to classify invertible fermionic phases with symmetry in terms of bosonic symmetry-enriched topological orders \cite{Bhardwaj_2017, Barkeshli_2022}. Moreover, the results of computing the higher dimensional classifications of bosonic and fermionic SPT protected by ordinary symmetries do not bear any obvious relation \cite{Wang_2018,Wang_2020}, suggesting that a correspondence -- if it exists at all -- is fairly subtle. The existence (under certain conditions) of a stacking-compatible correspondence between bosonic and fermionic SPT phases may be a special property of bosonization in one dimension.

Where the isomorphism of SPT phases exists, we have written an explicit form of its action on supercohomology invariants, which we contrasted with that of the Jordan-Wigner transformation. We have then asked whether this new transformation could be thought of as a generalized ``stacking-compatible'' Jordan-Wigner transformation. In short, the answer is no, the reason being that the isomorphism acts on the set of SPT phases in a way that a true duality cannot. This result highlights that, even when the bosonic and fermionic classifications are isomorphic, there are fundamental differences between bosonic and fermionic systems that cannot be compensated for by duality. While we treated duality in the abstract language of symmetry TFT, we expect this answer means there is no transformation from spins to fermions on the lattice (like the Jordan-Wigner transformation) that is compatible with the bosonic and fermionic SPT stacking rules. It would be interesting to test this expectation by working directly with lattice Hamiltonian models of SPT phases. It would also be interesting to generalize this study to bosonic and fermionic phases protected by generalized symmetries and their stacking structures. We note that, while dualities between two given one dimensional systems had been extensively studied from the symmetry TFT perspective in prior work, our investigation has been concerned with the action of dualities on the entire collection of phases and the relation of dualities to the stacking structure on this collection. There remains the question of how to physically interpret this nonduality -- the isomorphism of stacking rules that is not a consequence of duality -- and it is curious that antiunitary symmetry is capable of spoiling the isomorphism while unitary symmetry is not.

Along the way, we have generated several smaller results which readers may find interesting independently of the main results. We have written explicit $\cG$-pin partition functions for fermionic SPT phases, in terms of which we have rederived the stacking rule of invertible fermionic phases and defined a nonsplit version of the Jordan-Wigner transformation; we have given a complete gauge-invariant characterization of degree two twisted cohomology classes in terms of bosonic partition functions on genus $n$ orientable and nonorientable surfaces; we have studied when twisted cohomology classes are of order two and have found examples where they are not; we have discovered dualities that act on the group of SPT phases but not by automorphisms. Many of these results are contained in the following appendices.

\appendix

\section{Invertible fermionic phases}\label{sec:invertible}

The SPT phases discussed so far are examples of \emph{invertible phases}, i.e. phases with an inverse under stacking. When the symmetry group splits as $\cG=G_b\times\ZZ_2^f$, there are invertible phases beyond SPT phases. Rather than becoming trivial upon forgetting the symmetry (as SPT phases do), these additional invertible phases reduce to the nontrivial invertible topological order. This topological order is realized on the lattice as a phase of Majorana chains \cite{kitaev2001,FidkowskiKitaev} and has the Arf invariant as its topological partition function \cite{kttw}; the full Arf topological field theory has also been described \cite{Debray_2018,Turzillo_2020,Kobayashi_2019}. In total, invertible fermionic phases are classified by three invariants:\footnote{When $\gamma=1$, the invariant $\nu$ is defined differently from Eq. \eqref{nu} and satisfies $\nu(p)=1$ instead of $\nu(p)=0$ \cite{FidkowskiKitaev}; therefore, the pullback $\beta=s^*\nu$ \eqref{pullbacktoGb} satisfies $\delta\beta=\gamma\rho$. Upon choosing a splitting, the usual $\delta\beta=0$ is recovered. But if the group does not split, no $\beta$ is compatible with $\gamma=1$ (i.e. $\gamma=1$ is invalid), so no additional phases beyond SPT phases are possible \cite{FidkowskiKitaev}.} the usual $\alpha$ and $\beta$, and a new invariant
\begin{equation}
    \gamma\in H^0(G_b,\ZZ_2)=\ZZ_2
\end{equation}
that encodes whether the phase is SPT ($\gamma=0$) or not ($\gamma=1$). The most general invertible fermionic theory with split symmetry has partition function
\begin{equation}
    \cZ_{\alpha,\beta,\gamma}^f[X,A_b,\eta]=\exp(2\pi i\int_XA_b^*\alpha)\exp(\frac{\pi i}{2}q_{\eta}(A_b^*\beta))\text{Arf}(\eta)^\gamma~,
\end{equation}
which comes from stacking the theory \eqref{fermionictheorysplit} with the invertible topological order $\text{Arf}(\eta)^\gamma$.

As before, the stacking rule is obtained by multiplying partition functions:
\begin{align}\begin{split}
    \cZ_{\alpha_1,\beta_1,\gamma_1}^f[X,A_b,\eta]&\cZ_{\alpha_2,\beta_2,\gamma_2}^f[X,A_b,\eta]\\
    &=\exp(2\pi i\int_XA_b^*(\alpha_1+\alpha_2))\exp(\frac{\pi i}{2}(q_{\eta}(A_b^*\beta_1)+q_\eta(A_b^*\beta_2)))\text{Arf}(\eta)^{\gamma_1+\gamma_2}\\
    &=\exp(2\pi i\int_XA_b^*(\alpha_1+\alpha_2+\tfrac{1}{2}\beta_1\cup\beta_2))\\
    &\qquad\qquad\qquad\qquad\times\,\exp(\frac{\pi i}{2}(q_{\eta}(A_b^*(\beta_1+\beta_2))+\gamma_1\gamma_2q_\eta(A_b^*x)))\text{Arf}(\eta)^{\gamma_1+\gamma_2\text{ mod }2}\\
    &=\exp(2\pi i\int_XA_b^*(\alpha_1+\alpha_2+\tfrac{1}{2}\beta_1\cup\beta_2+\tfrac{1}{2}\gamma_1\gamma_2(\beta_1+\beta_2)\cup x))\\
    &\qquad\qquad\qquad\qquad\times\,\exp(\frac{\pi i}{2}q_{\eta}(A_b^*(\beta_1+\beta_2+x)))\text{Arf}(\eta)^{\gamma_1+\gamma_2\text{ mod }2}\\
    &=\cZ_{\alpha_1+\alpha_2+\tfrac{1}{2}\beta_1\cup\beta_2+\tfrac{1}{2}\gamma_1\gamma_2(\beta_1+\beta_2)\cup x,\beta_1+\beta_2+x,\beta_1+\beta_2}^f[X,A_b,\eta]~,
\end{split}\end{align}
where in the second equality we used the relation \eqref{arfsquared} and the constraint $A_b^*x=w_1$ to absorb $\text{Arf}(\eta)^2$ into the middle term. From this calculation, the full stacking rule may be read off as
\begin{equation}\label{fullstacking}
    \left[\begin{array}{c}\alpha_1\\\beta_1\\\gamma_1\end{array}\right]\otimes_\cF\left[\begin{array}{c}\alpha_2\\\beta_2\\\gamma_2\end{array}\right]=\left[\begin{array}{c}\alpha_1+\alpha_2+\tfrac{1}{2}\beta_1\beta_2+\tfrac{1}{2}\gamma_1\gamma_2(\beta_1+\beta_2)x\\\beta_1+\beta_2+\gamma_1\gamma_2x\\\gamma_1+\gamma_2\end{array}\right]~.
\end{equation}
If restricted to SPT phases ($\gamma=0$), this expression matches the SPT stacking rule \eqref{fermionicstacking}. Also note that if the symmetry is unitary ($x=0$), the stacking rule splits as $\text{SPTs}\times\ZZ_2$.

The stacking rule was first stated for split, unitary symmetry groups in Ref. \cite{GK} and derived for arbitrary unitary symmetries in Refs. \cite{Bultinck_2017,KTY16}, where it takes the form \eqref{fullstacking} with $x=0$. The stacking rule for invertible fermionic phases with general symmetry first appeared in Ref. \cite{Bultinck_2017}, where it takes the form\footnote{In the stacking rule for $\gamma_1=\gamma_2=1$ (Eq. 132) of Ref. \cite{Bultinck_2017}, the factor of $\tfrac{1}{2}\tilde\beta_1(g)(\tilde\beta_2(h)+x(h))+\tfrac{1}{4}(\tilde\beta_2(g)+x(g)+\tilde\beta_2(h)+x(h)-\tilde\beta_2(gh)-x(gh))$ can be rewritten as $\tfrac{1}{2}\tilde\beta_1\cup(\tilde\beta_2+x)+\tfrac{1}{2}(\tilde\beta_2+x)^2$, which is equal to $\tfrac{1}{2}\tilde\beta_1\tilde\beta_2+(\tilde\beta_1+\tilde\beta_2+x)x$.} (in our notation)
\begin{equation}\label{nickstacking}
    \left[\begin{array}{c}\tilde\alpha_1\\\tilde\beta_1\\\gamma_1\end{array}\right]\otimes_\cF\left[\begin{array}{c}\tilde\alpha_2\\\tilde\beta_2\\\gamma_2\end{array}\right]=\left[\begin{array}{c}\tilde\alpha_1+\tilde\alpha_2+\tfrac{1}{2}\tilde\beta_1\tilde\beta_2+\tfrac{1}{2}\gamma_1\gamma_2(\tilde\beta_1+\tilde\beta_2+x)x\\\tilde\beta_1+\tilde\beta_2+\gamma_1\gamma_2x\\\gamma_1+\gamma_2\end{array}\right]~.
\end{equation}
The extra factor of $\tfrac{1}{2}\gamma_1\gamma_2x^2$ relative to the rule \eqref{fullstacking} can be absorbed by the following redefinition:
\begin{equation}\label{alphashift}
    \tilde\alpha = \alpha + \tfrac{1}{2}\gamma\beta x
\end{equation}
\begin{equation}\label{betashift}
    \tilde\beta = \beta + \gamma x~,
\end{equation}
which shifts the $\alpha$ stacking rule by $\tfrac{1}{2}\gamma_1\gamma_2x^2$, precisely the difference between the expressions. Also note that, under the redefinition \eqref{betashift} of $\beta$ alone, the $\alpha$ rule shifts by $\tfrac{1}{2}\gamma_1\beta_2x+\tfrac{1}{2}\gamma_2\beta_1x+\tfrac{1}{2}\gamma_1\gamma_2x^2$, resulting in a rule
\begin{equation}\label{oldstacking}
    \left[\begin{array}{c}\tilde\alpha_1\\\beta_1\\\gamma_1\end{array}\right]\otimes_\cF\left[\begin{array}{c}\tilde\alpha_2\\\beta_2\\\gamma_2\end{array}\right]=\left[\begin{array}{c}\tilde\alpha_1+\tilde\alpha_2+\tfrac{1}{2}\beta_1\beta_2+\tfrac{1}{2}(1-\gamma_1)\gamma_2\beta_1x+\tfrac{1}{2}(1-\gamma_2)\gamma_1\beta_2x\\\beta_1+\beta_2+\gamma_1\gamma_2x\\\gamma_1+\gamma_2\end{array}\right]~.
\end{equation}
This form of the rule was first derived in Ref. \cite{TY17} and reproduced in Refs. \cite{Bourne,Aksoy}. It is in these variables that the $\ZZ_8$ stacking rule for time-reversal symmetric Majorana chains was first expressed \cite{FidkowskiKitaev}. A further redefinition \eqref{alphashift} in $\alpha$ shifts the $\alpha$ rule by $\tfrac{1}{2}\gamma_1\beta_1x+\tfrac{1}{2}\gamma_2\beta_2x-\tfrac{1}{2}(\gamma_1+\gamma_2)(\beta_1+\beta_2+\gamma_1\gamma_2x)x=\tfrac{1}{2}\gamma_1\beta_2x+\tfrac{1}{2}\gamma_2\beta_1x$, bringing it into the form \eqref{fullstacking} derived by stacking partition functions.

This differences between the formulations \eqref{fullstacking}, \eqref{nickstacking}, and \eqref{oldstacking} of the stacking rule have the following interpretations. In a system with $\gamma=1$, the symmetry $G_b\times\ZZ_2$ acts on an algebra of local operators $\End(U_b)\otimes C\ell(1)$ in a manner specified by the invariants $[\alpha,\beta]$. The invariant $\beta$ measures whether a symmetry commutes or anticommutes with the generator $\Gamma$ of $\CC\ell(1)$ squaring to $+1$ \cite{FidkowskiKitaev,TY17}, whereas $\tilde\beta$ measures commutation with $i\Gamma$ \cite{Bultinck_2017}; for this reason, $\beta$ and $\tilde\beta$ differ by $x$. The difference between $\alpha$ and $\tilde\alpha$ has to do with a choice in boundary condition. Following Ref. \cite{susy} (see arXiv v2), a boundary condition amounts to a choice $\delta\in H^1(G_b,\ZZ_2)$ of compatible symmetry action on the module $U_b\otimes\CC^{1|1}$ over the algebra:
\begin{equation}
    g_b\mapsto Q(g_b)\otimes X^{\delta(g_b)}Z^{\beta(g_b)}~,
\end{equation}
where $Q$ is a projective representation of $G_b$ of class $[\tilde\alpha]$. In total, the projectivity class of this action is
\begin{equation}
    \tilde\alpha+\tfrac{1}{2}\beta\cup\delta~.
\end{equation}
The choice $\delta=0$ is often made implicitly and realizes $\tilde\alpha$ as the projectivity in the boundary symmetry action. On the other hand, the choice $\delta=x$ corresponds to $\alpha$. The formulas \eqref{fullstacking} and \eqref{oldstacking} are the stacking rule written in variables where the $2$-cocycle is the one measuring the boundary projectivity -- either $\alpha$ or $\tilde\alpha$. For example, in the case of symmetry $\cG=\ZZ_2^T\times\ZZ_2^f$, the three invariants $[\alpha],[\beta],[\gamma]\in\ZZ_2$ form the classification group $\ZZ_8$ according to the binary expansion \cite{Delmastro_2021}, whereas $[\tilde\alpha],[\beta],[\gamma]$ form the $\ZZ_8$ a different way \cite{FidkowskiKitaev,TY17}. This difference appears in whether, for phase $3\in\ZZ_8$, the boundary action of time-reversal symmetry squares to $+1$ (binary) or $-1$ (usual), which is a choice of boundary condition independent of the topological invariants.

Finally, we remark that passing an odd ($\gamma=1$) fermionic phase through the inverse transformation yields
\begin{align}\begin{split}
    \cJ\cW^{-1}&(\cZ^f_{\alpha,\beta,1})[X,A_b,A_p]\\
    &=\sum_\eta\exp(2\pi i\int_XA_b^*\alpha)\exp(\frac{\pi i}{2}q_{\eta}(A_b^*\beta))\text{Arf}(\eta)\exp(\frac{\pi i}{2}q_{\eta+\tau}(A_p))\frac{\text{Arf}(\eta)^{-1}}{\sqrt{|H^1(X,\ZZ_2)|}}\\
    &=\frac{1}{\sqrt{|H^1(X,\ZZ_2)|}}\exp(2\pi i\int_XA_b^*\alpha+\tfrac{1}{2}A_b^*\beta\cup A_p)\sum_\eta\exp(\frac{\pi i}{2}q_{\eta}(A_b^*\beta+A_p))\\
    &=\sqrt{|H^1(X,\ZZ_2)|}\,\delta(A_b^*\beta+A_p)\,\cZ^b_{\alpha,\beta}[X,A_b,A_p]~,
\end{split}\end{align}
where the rule \eqref{quadratic} was used in the second line and the relation \eqref{sumsigma} in the third. The $\delta$-function constraint on the $\ZZ_2^f$ background gauge field $A_p$ indicates that the $\ZZ_2^f$ symmetry is broken, with the pattern of symmetry breaking depending on the topological invariant $\beta$.


\section{State sum evaluation of bosonic partition functions}\label{sec:statesum}

To compute the evaluations \eqref{orientableevaluated} and \eqref{nonorientableevaluated} of partition functions $\cZ_\omega^b[X,\cA]$ stated in \cref{sec:genusrules}, we use the state sum construction \cite{Fukuma_1994,turaev2010homotopy,DijkgraafWitten,ShiozakiRyu,KTYbosonic}. The idea is to define the partition function on a discretization of the space $(X,\cA)$ and then check that the result is independent of the choice of discretization.

\begin{figure}
\centering
\begin{tikzpicture}[scale=2]

\draw[decoration={markings, mark=at position 0.18 with {\arrow[scale=2]{<}}}, decoration={markings, mark=at position 0.51 with {\arrow[scale=2]{>}}}, decoration={markings, mark=at position 0.84 with {\arrow[scale=2]{>}}}, postaction={decorate}] (0,0) -- (-1,-1.73) node[midway,left]{$gh$\,\,\,} -- (1,-1.73) node[midway,below]{\,\,\,\,$g$} -- (0,0) node[midway,right]{\,\,\,$h$};

\node at (0,-1.1) {$+$};

\begin{scope}[xshift=75]

\draw[decoration={markings, mark=at position 0.07 with {\arrow[scale=1]{<}}}, decoration={markings, mark=at position 0.20 with {\arrow[scale=1]{<}}}, decoration={markings, mark=at position 0.32 with {\arrow[scale=1]{>}}}, decoration={markings, mark=at position 0.45 with {\arrow[scale=1]{>}}}, decoration={markings, mark=at position 0.57 with {\arrow[scale=1]{<}}}, decoration={markings, mark=at position 0.70 with {\arrow[scale=1]{<}}}, decoration={markings, mark=at position 0.82 with {\arrow[scale=1]{>}}}, decoration={markings, mark=at position 0.94 with {\arrow[scale=1]{>}}}, postaction={decorate}] (0,0) -- (-0.5,-0.5) node[midway,left]{$g_2$\,} -- (-0.5,-1.21) node[midway,left]{$h_2$\,} -- (0,-1.71) node[midway,left]{$g_1$\,} -- (0.71,-1.71) node[midway,below]{$h_1$\,} -- (1.21,-1.21) node[midway,right]{\,$g_1$} -- (1.21,-0.5) node[midway,right]{\,$h_1$} -- (0.71,0) node[midway,right]{\,$g_2$} -- (0,0) node[midway,above]{$h_2$\,};

\draw[decoration={markings, mark=at position 0.52 with {\arrow[scale=1]{>}}}, postaction={decorate}] (-0.5,-1.21) -- (0,0);
\draw[decoration={markings, mark=at position 0.52 with {\arrow[scale=1]{>}}}, postaction={decorate}] (-0.5,-1.21) -- (0.71,0);
\draw[decoration={markings, mark=at position 0.52 with {\arrow[scale=1]{>}}}, postaction={decorate}] (-0.5,-1.21) -- (1.21,-0.5);
\draw[decoration={markings, mark=at position 0.52 with {\arrow[scale=1]{>}}}, postaction={decorate}] (-0.5,-1.21) -- (1.21,-1.21) node[midway,above]{\small\qquad$g_1h_1g_1^{-1}$};
\draw[decoration={markings, mark=at position 0.52 with {\arrow[scale=1]{>}}}, postaction={decorate}] (-0.5,-1.21) -- (0.71,-1.71) node[midway,right]{\small\,\,\,\,\,\,\,$g_1h_1$};

\node at (-0.4,-0.55) {$-$};
\node at (0.15,-0.28) {$+$};
\node at (0.65,-0.45) {$+$};
\node at (0.8,-0.9) {$-$};
\node at (0.68,-1.35) {$-$};
\node at (0.12,-1.6) {$+$};

\end{scope}

\begin{scope}[xshift=160]

\draw[decoration={markings, mark=at position 0.07 with {\arrow[scale=1]{>}}}, decoration={markings, mark=at position 0.20 with {\arrow[scale=1]{>}}}, decoration={markings, mark=at position 0.32 with {\arrow[scale=1]{>}}}, decoration={markings, mark=at position 0.45 with {\arrow[scale=1]{>}}}, decoration={markings, mark=at position 0.57 with {\arrow[scale=1]{>}}}, decoration={markings, mark=at position 0.70 with {\arrow[scale=1]{>}}}, decoration={markings, mark=at position 0.82 with {\arrow[scale=1]{>}}}, decoration={markings, mark=at position 0.94 with {\arrow[scale=1]{>}}}, postaction={decorate}] (0,0) -- (-0.5,-0.5) node[midway,left]{$g_4$\,} -- (-0.5,-1.21) node[midway,left]{$g_4$\,} -- (0,-1.71) node[midway,left]{$g_1$\,} -- (0.71,-1.71) node[midway,below]{$g_1$\,} -- (1.21,-1.21) node[midway,right]{\,$g_2$} -- (1.21,-0.5) node[midway,right]{\,$g_2$} -- (0.71,0) node[midway,right]{\,$g_3$} -- (0,0) node[midway,above]{$g_3$\,};

\draw[decoration={markings, mark=at position 0.52 with {\arrow[scale=1]{>}}}, postaction={decorate}] (-0.5,-1.21) -- (0,0);
\draw[decoration={markings, mark=at position 0.52 with {\arrow[scale=1]{>}}}, postaction={decorate}] (-0.5,-1.21) -- (0.71,0);
\draw[decoration={markings, mark=at position 0.52 with {\arrow[scale=1]{>}}}, postaction={decorate}] (-0.5,-1.21) -- (1.21,-0.5);
\draw[decoration={markings, mark=at position 0.52 with {\arrow[scale=1]{>}}}, postaction={decorate}] (-0.5,-1.21) -- (1.21,-1.21) node[midway,above]{\small\qquad$g_1^2g_2$};
\draw[decoration={markings, mark=at position 0.52 with {\arrow[scale=1]{>}}}, postaction={decorate}] (-0.5,-1.21) -- (0.71,-1.71) node[midway,right]{\small\,\,\,\,\,\,\,$g_1^2$};

\node at (-0.4,-0.55) {$+$};
\node at (0.15,-0.28) {$+$};
\node at (0.65,-0.45) {$+$};
\node at (0.8,-0.9) {$+$};
\node at (0.68,-1.35) {$+$};
\node at (0.12,-1.6) {$+$};

\end{scope}

\end{tikzpicture}
\caption{(left) A triangle that makes a contribution \eqref{statesumcontrib} to the state sum. (center) A discretization of the orientable surface of genus $n=2$. (right) A discretization of the nonorientable surface of genus $n=4$.}
\label{fig:statesum}
\end{figure}
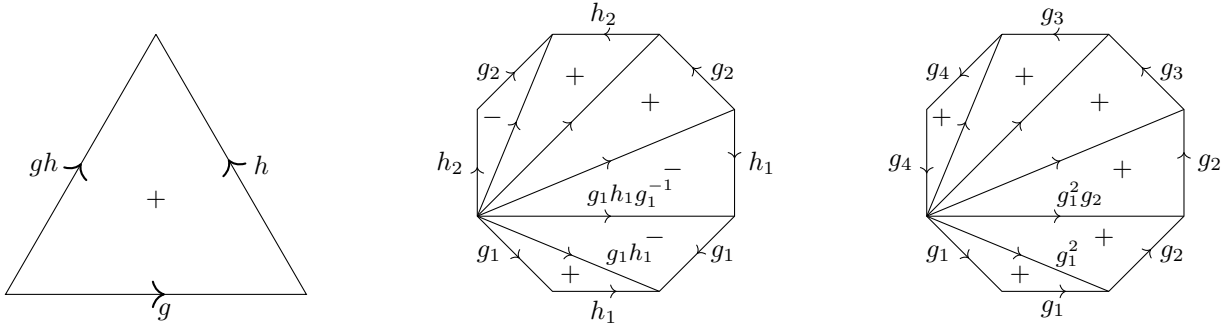

For theories with no antiunitary symmetries, $x=0$ -- where only orientable $X$ appear due to the constraint $\cA^*x=w_1$ -- a discretization consists of a triangulation of $X$, a branching structure on the edges of the triangulation (i.e. an assignment of edge directions such that there are no closed loops), group labels $g\in\cG$ on the edges such that the product of group labels (with labels inverted when the edge directions point against the path direction) around each cycle in the edges of $X$ equals the holonomy of $\cA$ along that cycle, and an orientation. The partition function is defined on such a discretization as the product of contributions
\begin{equation}\label{statesumcontrib}
    \cZ^b_\omega(\Delta_{g,h}^\pm)=\exp(\pm 2\pi i\,\omega(g,h))
\end{equation}
for each triangle whose two legs pointing in the same direction are labeled by $g$ and $h$, as depicted in \cref{fig:statesum}; the triangle enters with a $+$ sign ($-$ sign) if its local orientation defined by the branching structure agrees (disagrees) with the global orientation. One can check that independence of this partition function on the discretization -- that is, invariance under retriangulation by Pachner moves, change of the branching structure, and homotopy of the group labels -- is all guaranteed by the cocycle condition on $\omega$. The generalization of this construction to general symmetries $(G,x)$ has some subtleties, which we will not explore here; nevertheless, the partition functions on nonorientable surfaces can be recovered from the discretizations of \cref{fig:statesum}, where the triangles of the nonorientable surfaces all enter with $+$ signs.

First consider the orientable surface of genus $n$ labeled by holonomies $\{g_i,h_i\}$ satisfying $x(g_i)=x(h_i)=0$ and $\prod_ig_ih_ig_i^{-1}h_i^{-1}=1$. A discretization of this space by $4n-2$ triangles is depicted in \cref{fig:statesum}. On this discretization, the state sum evaluates to
\begin{equation}
    \cZ^b_\omega[(T^2)^{\#n},\{g_i,h_i\}]=\exp(2\pi i\sum_{i=1}^n \omega(x_i,g_i)+\omega(x_ig_i,h_i)-\omega(x_ig_ih_ig_i^{-1},g_i)-\omega(x_ig_ih_ig_i^{-1}h_i^{-1},h_i))~,
\end{equation}
where again $x_i=\prod_{j<i}g_jh_jg_j^{-1}h_j^{-1}$. Since $x_1=x_ng_nh_ng_n^{-1}h_n^{-1}=1$, the expression contains only $4n-2$ terms. Using the cocycle conditions
\begin{equation}
    -\omega(x_ig_ih_ig_i^{-1},g_i)=\omega(x_ig_ih_i,g_i^{-1})-\omega(g_i^{-1},g_i)~,
\end{equation}
\begin{equation}
    -\omega(x_ig_ih_ig_i^{-1}h_i^{-1},h_i)=\omega(x_ig_ih_ig_i^{-1},h_i^{-1})-\omega(h_i^{-1},h_i)~,
\end{equation}
we obtain the form \eqref{orientableevaluated} stated in \cref{sec:genusrules}. The check of gauge invariance performed in \cref{sec:genusrules} suggests that this answer is correct. But as one extra reality check, one can compute the torus ($n=1$, $ghg^{-1}=h$)
\begin{equation}\label{torus}
    \cZ^b_\omega[T^2_{g,h}]=\exp(2\pi i\,(\omega(g,h)-\omega(h,g)))
\end{equation}
and see that it matches the known result \cite{ShiozakiRyu,KTYbosonic}. 

Next consider the nonorientable surface of genus $n$ labeled by holonomies $\{g_i\}$ satisfying $x(g_i)=1$ and $\prod_i g_i^2=1$. A discretization of this space is depicted in \cref{fig:statesum}. We obtain the expression
\begin{equation}
    \cZ^b_\omega[(\RR P^2)^{\#n},\{g_i\}]=\exp(2\pi i\sum_{i=1}^n \omega(y_i,g_i)+\omega(y_ig_i,g_i))~,
\end{equation}
where again $y_i=\prod_{j<i}g_j^2$. This form is related to the expression \eqref{nonorientableevaluated} stated in \cref{sec:genusrules} by the cocycle condition
\begin{equation}
    \omega(y_i,g_i)+\omega(y_ig_i,g_i)=\omega(g_i,g_i)+\omega(y_i,g_i^2)~.
\end{equation}
Computing $\RR P^2$ ($n=1$, $g^2=1$)
\begin{equation}
    \cZ^b_\omega[\RR P^2_g]=\exp(2\pi i\,\omega(g,g))~,
\end{equation}
the known result is recovered \cite{ShiozakiRyu}. Now test it on the Klein bottle ($n=2$, $g^2h=h^{-1}$):
\begin{equation}\label{klein}
    \cZ^b_\omega[K_{g,h}]=\exp(2\pi i\,(\omega(g,g)+\omega(g^2,h)+\omega(h^{-1},h)))~.
\end{equation}
To compare to Ref. \cite{ShiozakiRyu}, substitute $h\mapsto g^{-1}k^{-1}$ in the expression. The new $k$ has $x(k)=0$ and $gk^{-1}=kg$, and the expression becomes
\begin{equation}
    \cZ^b_\omega[K_{g,g^{-1}k^{-1}}]=\exp(2\pi i\,(\omega(g,g)+\omega(g^2,g^{-1}k^{-1})+\omega(kg,g^{-1}k^{-1})))~.
\end{equation}
Use the twisted cocycle conditions
\begin{equation}
    \omega(g,g)+\omega(g^2,g^{-1}k^{-1})=\omega(g,k^{-1})-\omega(g,g^{-1}k^{-1})~,
\end{equation}
\begin{equation}
    \omega(kg,g^{-1}k^{-1})=\omega(k,k^{-1})-\omega(k,g)+\omega(g,g^{-1}k^{-1})
\end{equation}
to obtain
\begin{equation}
    \cZ^b_\omega[K_{g,g^{-1}k^{-1}}]=\exp(2\pi i\,(\omega(g,k^{-1})+\omega(k,k^{-1})-\omega(k,g)))~,
\end{equation}
which matches the expression of Ref. \cite{ShiozakiRyu}.

\section{Twisted cohomology not of order two}\label{sec:ordertwo}

In this section, we argue that the presence of an antiunitary symmetry in the center of $\cG$ implies that the twisted cohomology classes of $\cG$ are all of order two:
\begin{equation}
    2[\omega]=0~,\qquad[\omega]\in H^2(\cG,U(1)_x)~.
\end{equation}
We also provide examples of classes \emph{not} of order two that arise when no antiunitary symmetry lies in the center, and we discuss how they spoil the converse to \hyperref[lemma3]{Lemma 3}. These examples are surprising, as the groups with antiunitary symmetry that are common in the physics literature have classes only of order two (see for example the table in Ref. \cite{CGLW}) and one might have expected this to be a general rule \cite{mathoverflow}.

Conjugating the arguments of a cocycle is a gauge transformation up to a sign, as we can see by applying the cocycle conditions on the triplets $(k,k^{-1}gk,k^{-1}hk)$, $(g,k,k^{-1}hk)$, and $(g,h,k)$ to obtain
\begin{align}\begin{split}\label{conjgauge}
    (-1)^{x(k)}\omega(k^{-1}gk,k^{-1}hk)
    &=\omega(k,k^{-1}gk)-\omega(k,k^{-1}ghk)+\omega(gk,k^{-1}hk)\\
    &=\omega(k,k^{-1}gk)-\omega(k,k^{-1}ghk)+(-1)^{x(g)}\omega(k,k^{-1}hk)-\omega(g,k)+\omega(g,hk)\\
    &=\omega(k,k^{-1}gk)-\omega(k,k^{-1}ghk)+(-1)^{x(g)}\omega(k,k^{-1}hk)-\omega(g,k)\\
    &\qquad\qquad\qquad\qquad\qquad\qquad\qquad\quad\,\,\,+\omega(gh,k)-(-1)^{x(g)}\omega(h,k)+\omega(g,h)\\
    &=\omega(g,h)+(\delta\lambda)(g,h)~,
\end{split}\end{align}
where $\lambda(g)=(i_g\omega)(k)=\omega(k,k^{-1}gk)-\omega(g,k)$. If $k$ is antiunitary and commutes with $g$ and $h$, this reads
\begin{equation}
    2\omega(g,h)=(\delta\lambda)(g,h)~.
\end{equation}
Therefore, when this holds for all $g,h$ (i.e. when $k$ is central), we have $2[\omega]=0$, as claimed.

Now we give examples where $[\omega]$ is not of order two, despite $\cG$ having a (noncentral) antiunitary symmetry. As noted previously, the partition function of $\RR P^2$ is always of order two; however, the higher order of the class shows up in other partition functions -- both orientable and nonorientable -- that do not square to one.

A first example is the group
\begin{equation}
    \cG=\ZZ_3\times D_6^T~,
\end{equation}
where $x$ is $1$ on reflections in the dihedral group $D_6^T=\ZZ_3\rtimes\ZZ_2^T$. The group law is
\begin{equation}
    (a,b,c)\cdot(a',b',c')=(a+a',b+(-1)^cb',c+c')~.
\end{equation}
A projective representation of the group is given by
\begin{equation}
    Q(a,b,c)=X^aZ^b\cK^c~,
\end{equation}
where $X$ and $Z$ denote the ``shift matrix'' and ``clock matrix'' \cite{clockshift}
\begin{equation}
    X=\begin{pmatrix}0&0&1\\1&0&0\\0&1&0\end{pmatrix}~,\qquad    Z=\begin{pmatrix}1&0&0\\0&\omega_3&0\\0&0&\omega_3^2\end{pmatrix}~,
\end{equation}
where $\omega_3$ is a third root of unity, and $\cK$ denotes complex conjugation in this basis. These operations satisfy
\begin{equation}
    X^3=Z^3=\cK^2=1~,\qquad ZX=\omega_3XZ~,\qquad\cK X=X\cK~,\qquad\cK Z=Z^{-1}\cK~.
\end{equation}
Therefore
\begin{align}\begin{split}
    Q(a,b,c)Q(a',b',c')
    &=X^aZ^b\cK^cX^{a'}Z^{b'}\cK^{c'}\\
    &=X^aZ^bX^{a'}Z^{(-1)^cb'}\cK^{c+c'}\\
    &=\omega_3^{a'b}X^{a+a'}Z^{b+(-1)^cb'}\cK^{c+c'}\\
    &=\omega_3^{a'b}Q((a,b,c)\cdot(a',b',c'))~.
\end{split}\end{align}
The values
\begin{equation}
    \exp(2\pi i\,\omega((a,b,c),(a',b',c')))=\omega_3^{a'b}
\end{equation}
are guaranteed to satisfy the twisted cocycle condition by the associativity of multiplication in the projective representation, and they are of order three, not two. This cocycle defines torus partition functions that can be nontrivial third roots of unity, as the formula \eqref{torus} shows:
\begin{equation}
    \cZ^b_\omega[T^2,\{(a,b,0),(a',b',0)\}]=\exp(2\pi i\,(\omega((a,b,0),(a',b',0))-\omega((a',b',0),(a,b,0))))=\omega_3^{a'b-ab'}~,
\end{equation}
noting that arbitrary unitary symmetries commute and so respect the torus fundamental group. Klein bottle partition functions \eqref{klein} are defined for any $a+a'\in\{0,2\}$ since $1=g^2h^2=(2a+2a',0,0)$:
\begin{align}\begin{split}
    &\cZ^b_\omega[K,\{(a,b,1),(a',b',1)\}]\\
    &\qquad\qquad\qquad=\exp(2\pi i\,(\omega((a,b,1),(a,b,1))+\omega((2a,0,0),(a',b',1))+\omega((-a',b',1),(a',b',1))))\\
    &\qquad\qquad\qquad=\omega_3^{ab+a'b'}~,
\end{split}\end{align}
using $(a,b,c)^{-1}=(-a,(-1)^{c+1}b,c)$. We see they can also be nontrivial third roots of unity. This example generalizes straightforwardly to the symmetry $\cG=\ZZ_n\times D_{2n}^T$, which has a twisted cohomology class of order $n$ that gives a torus and Klein bottle partition functions in $n^\text{th}$ roots of unity.

A second example is the group
\begin{equation}
    \cG=\ZZ_4\rtimes\ZZ_4^T~,
\end{equation}
where $x:\ZZ_4^T\mapsto\ZZ_2^T$ is the nontrivial map. The group law is
\begin{equation}
    (a,b)\cdot(a',b')=(a+(-1)^ba',b+b')~.
\end{equation}
A projective representation of the group is given by
\begin{equation}
    Q(a,b)=Z^a(X\cK)^b~,
\end{equation}
where $X$ and $Z$ are the obvious rank $4$ generalizations of the shift and clock matrices above. Then observe
\begin{align}\begin{split}
    Q(a,b)Q(a',b')
    &=Z^aX^b\cK^bZ^{a'}X^{b'}\cK^{b'}\\
    &=Z^aX^bZ^{(-1)^ba'}X^{b'}\cK^{b+b'}\\
    &=\omega_4^{(-1)^ba'b}Z^{a+(-1)^ba'}X^{b+b'}\cK^{b+b'}\\
    &=\omega_4^{(-1)^ba'b}Q((a,b)\cdot(a',b'))
\end{split}\end{align}
The cocycle
\begin{equation}\label{examplecocycle}
    \exp(2\pi i\,\omega((a,b),(a',b')))=\omega_4^{(-1)^ba'b}
\end{equation}
has torus
\begin{equation}
    \cZ^b_\omega[T^2,\{(a,b),(a',b')\}]=\exp(2\pi i\,(\omega((a,b),(a',b'))-\omega((a',b'),(a,b))))=\omega_4^{a'b-ab'}~,
\end{equation}
defined for $b,b'=0,2$, and Klein bottle
\begin{align}\begin{split}
    \cZ^b_\omega[K,\{(a,b),(a',b')\}]&=\exp(2\pi i\,(\omega((a,b),(a,b))+\omega((0,2b),(a',b'))+\omega((a',-b'),(a',b'))))\\
    &=\omega_4^{-ab+2a'b+a'b'}\\
    &=\omega_4^{-ab-a'b'}~,
\end{split}\end{align}
defined for $b,b'=1,3$, where we used $(a,b)^{-1}=((-1)^{b+1}a,-b)$, then $2(b+b')=0$ in the last line. As in the other example, the Klein bottle can be any fourth root of unity; but this time, the torus takes values in $\pm 1$ due to $b,b'=0,2$. In fact, for any orientable surface, the partition function for $\omega$ is the same as that for the pullback of $\omega$ to $\ker x$. In this example, $\ker x=\ZZ_4\times\ZZ_2$ has cohomology group $\ZZ_2$, which means that all of its orientable partition functions will be of order two. The nonorientable partition functions are
\begin{equation}\label{examplenonor}
    \cZ^b_\omega[(\RR P^2)^{\#n},\{(a_i,b_i)\}]=\exp(2\pi i\sum_{i=1}^n(\omega((a_i,b_i),(a_i,b_i))+\omega((0,2(i-1)),(0,2b_i))))=\omega_4^{-\sum_{i=1}^na_ib_i}~,
\end{equation}
for $b_i=1,3$. Now we claim that this $\omega$ is a square root
\begin{equation}
    2[\omega]=[\tfrac{1}{2}\nu\cup\nu]
\end{equation}
of the half cup square\footnote{A projective representation with this cocycle is given by $Q(a,b)=Z_2^aX_2^a\cK^b$ with $X_2$ and $Z_2$ the Pauli matrices.} of
\begin{equation}\label{examplebeta}
    \nu(a,b)=a\text{ mod }2~.
\end{equation}
This is equivalent to the partition functions associated with $2\omega$ and $\tfrac{1}{2}\nu\cup\nu$ agreeing on every space. Since $\omega$ is order two on orientable surfaces, its square vanishes on them, consistent with the result for $\tfrac{1}{2}\nu\cup\nu$ \eqref{orientableonbub}. On nonorientable surfaces, the square of the expression \eqref{examplenonor} is
\begin{equation}
    \cZ^b_\omega[(\RR P^2)^{\#n},\{(a_i,b_i)\}]^2=(-1)^{-\sum_{i=1}^na_ib_i}=(-1)^{\sum_{i=1}^na_i}=(-1)^{\sum_{i=1}^n\beta(a_i,b_i)}~,
\end{equation}
also consistent with the result for $\tfrac{1}{2}\nu\cup\nu$ \eqref{nonorientableonbub}. A counterexample to the full converse of \hyperref[lemma3]{Lemma 3} is therefore obtained from the fermionic symmetry group
\begin{equation}
    \cG=(\ZZ_4\rtimes\ZZ_4^T)\times\ZZ_2^f
\end{equation}
by taking $\beta$ to be the $\nu$ \eqref{examplebeta} defined above and the square root $R_\beta$ to be the $\omega$ \eqref{examplecocycle}. The half cup square of this $\beta$ has a square root, despite failing the \hyperref[genusrule]{genus $n$ rule} for $n$ even, and -- since the other choices of $\beta$ for this group also have square roots (namely, vanishing ones)\footnote{The half cup square of $\beta(a,b)=b\text{ mod }2$ is trivial and unrelated to the nontrivial degree two cohomology class of $\ZZ_4^T$.} -- the SPT classifications are isomorphic.

\section{The Arf invariant}\label{sec:arf}

In this appendix, we review some basic properties of the Arf invariant.

The inverse is given by
\begin{equation}\label{invarf}
    \text{Arf}(e)^{-1}=\frac{1}{\sqrt{|H^1(X,\ZZ_2)|}}\sum_{a}\exp(\frac{-\pi i}{2}q_e(a))~,
\end{equation}
as we can verify by computing
\begin{align}\begin{split}
    \text{Arf}(e)\text{Arf}(e)^{-1}
    &=\frac{1}{|H^1(X,\ZZ_2)|}\sum_{a,a'}\exp(\frac{\pi i}{2}(q_e(a)-q_e(a')))\\
    &=\frac{1}{|H^1(X,\ZZ_2)|}\sum_{a''=a+a'}\exp(\frac{\pi i}{2}q_e(a''))\sum_{a}\exp(\pi i\int_X a\cup a'')\\
    &=\sum_{a''}\exp(\frac{\pi i}{2}q_e(a''))\delta(a'')\\
    &=1~,
\end{split}\end{align}
where we have used the relation
\begin{equation}\label{delta}
    \sum_a\exp(\pi i\int_X a\cup b)=|H^1(X,\ZZ_2)|\,\delta(b)~.
\end{equation}
The square is given by
\begin{align}\begin{split}\label{arfsquared}
    \text{Arf}(e)^2
    &=\frac{1}{|H^1(X,\ZZ_2)|}\sum_{a,a'}\exp(\frac{\pi i}{2}(q_e(a)+q_e(a')))\\
    &=\frac{1}{|H^1(X,\ZZ_2)|}\sum_{a''=a+a'}\exp(\frac{\pi i}{2}q_e(a''))\sum_{a}\exp(\pi i\int_X a\cup (a+a''))\\
    &=\sum_{a''}\exp(\frac{\pi i}{2}q_e(a''))\delta(a''+w_1)\\
    &=\exp(\frac{\pi i}{2}q_e(w_1))~,
\end{split}\end{align}
where we have used the relation
\begin{equation}\label{delta-x}
    \sum_a\exp(\pi i\int_X a\cup (a+b))=|H^1(X,\ZZ_2)|\,\delta(b+w_1)~,
\end{equation}
which in turn comes from the Wu relation $w_1\cup a=a\cup a$ \cite{charclass}. It follows that the Arf invariant is an eighth root of unitary and that its fourth power is independent of the pin structure since
\begin{equation}
    \text{Arf}(e)^4=\exp(\frac{\pi i}{2}(q_e(w_1)+q_e(w_1)))=\exp(\pi i\int_Xw_1^2)\in\pm 1~.
\end{equation}

Summing over pin structures yields
\begin{align}\begin{split}\label{sumsigma}
    \sum_e\exp(\frac{\pi i}{2}q_e(a))
    &=\sum_b\exp(\frac{\pi i}{2}q_{e_0+b}(a))\\
    &=\exp(\frac{\pi i}{2}q_{e_0}(a))\sum_b\exp(\pi i\int_X a\cup b)\\
    &=\exp(\frac{\pi i}{2}q_{e_0}(a))|H^1(X,\ZZ_2)|\,\delta(a)\\
    &=|H^1(X,\ZZ_2)|\,\delta(a)~,
\end{split}\end{align}
using rule \eqref{shiftrule}. Therefore
\begin{equation}\label{sumarf}
    \frac{1}{\sqrt{|H^1(X,\ZZ_2)|}}\sum_e\text{Arf}(e)=\frac{1}{|H^1(X,\ZZ_2)|}\sum_{a,e}\exp(\frac{\pi i}{2}q_e(a))=1~.
\end{equation}

Finally, we have the relation
\begin{align}\begin{split}\label{qtimesarf}
    \exp(\frac{\pi i}{2}q_e(a))\text{Arf}(e+a)
    &=\frac{1}{\sqrt{|H^1(X,\ZZ_2)|}}\sum_b\exp(\frac{\pi i}{2}(q_e(a)+q_{e+a}(b)))\\
    &=\frac{1}{\sqrt{|H^1(X,\ZZ_2)|}}\sum_b\exp(\frac{\pi i}{2}(q_e(a)+q_e(b)))\exp(\pi i\int_Xa\cup b)\\
    &=\frac{1}{\sqrt{|H^1(X,\ZZ_2)|}}\sum_b\exp(\frac{\pi i}{2}q_e(a+b))\\
    &=\text{Arf}(e)~.
\end{split}\end{align}

We now return to the Jordan-Wigner transformation \eqref{jordanwigner}. The expression \eqref{jwinverse} is indeed the inverse:
\begin{align}\begin{split}
    \cJ\cW^{-1}\cJ\cW(\cZ^b)[X,A_b,A_p']
    &=\sum_{A_p,\eta}\frac{\text{Arf}(\eta+A_p)\text{Arf}(\eta+A_p')^{-1}}{|H^1(X,\ZZ_2)|}\cZ^b[X,A_b,A_p]\\
    &=\frac{1}{|H^1(X,\ZZ_2)|}\sum_{A_p,\eta}\exp(\frac{\pi i}{2}q_{\eta+A_p}(A_p+A_p'))\cZ^b[X,A_b,A_p]\\
    &=\sum_{A_p}\delta(A_p+A_p')\,\cZ^b[X,A_b,A_p]\\
    &=\cZ^b[X,A_b,A_p']~,
\end{split}\end{align}
by the rules \eqref{qtimesarf} and \eqref{sumsigma}. Alternatively, write the inverse as
\begin{equation}
    \cJ\cW^{-1}(\cZ^f)[X,A_b,A_p]=\sum_\eta\cZ^f[X,A_b,\eta]\exp(\frac{\pi i}{2}q_{\eta+\tau}(A_p+\tau))\frac{\text{Arf}(\eta+\tau)^{-1}}{\sqrt{|H^1(X,\ZZ_2)|}}~.
\end{equation}
and check
\begin{align}\begin{split}
    \cJ\cW^{-1}&(\cZ^f_{\alpha,\beta})[X,A_b,A_p]\\
    &=\sum_\eta\exp(2\pi i\int_XA_b^*\alpha+\tfrac{1}{2}A_b^*\beta\cup\tau)\exp(\frac{\pi i}{2}q_{\eta+\tau}(A_b^*\beta))\exp(\frac{\pi i}{2}q_{\eta+\tau}(A_p+\tau))\frac{\text{Arf}(\eta+\tau)^{-1}}{\sqrt{|H^1(X,\ZZ_2)|}}\\
    &=\exp(2\pi i\int_XA_b^*\alpha+\tfrac{1}{2}A_b^*\beta\cup A_p)\sum_\eta\exp(\frac{\pi i}{2}q_{\eta+\tau}(A_b^*\beta+A_p+\tau))\frac{\text{Arf}(\eta+\tau)^{-1}}{\sqrt{|H^1(X,\ZZ_2)|}}\\
    &=\exp(2\pi i\int_XA_b^*\alpha+\tfrac{1}{2}A_b^*\beta\cup A_p)\sum_\eta\frac{\text{Arf}(\eta+A_b^*\beta+A_p)^{-1}}{\sqrt{|H^1(X,\ZZ_2)|}}\\
    &=\cZ^b_{\alpha,\beta}[X,A_b,A_p]~.
\end{split}\end{align}
Here we used the freedom in the auxiliary variables to set the $\tau$ of $\cZ^f_{\alpha,\beta}$ equal to that of $\cJ\cW^{-1}$, used the relation \eqref{qtimesarf} in the third line, and used the relation \eqref{sumarf} in the last.

\section{Dualities of SPT phases: examples and a theorem}\label{sec:dualities_app}

The goal of this appendix is to prove the theorem stated in \cref{sec:dualities_Lag_SPT}. We first build intuition by analyzing several examples of dualities that map SPT phases to SPT phases, i.e. that do not induce symmetry-breaking in any SPT phase.\footnote{If one only wishes to map \emph{some} SPT phase to \emph{some other} SPT phase while not necessarily stabilizing the set of SPT phases, more general autoequivalence are allowed; these will induce symmetry breaking in \emph{other} SPT phases.} Then we ask when such dualities can implement a stacking isomorphism.

First consider $\cG=\ZZ_2$. The fusion ring has an automorphism that fixes $1,m$ and exchanges $e,em$; this preserves SPT phases (the trivial phase consisting of $1,m$) but not pure charges. But this map is not induced by a braided autoequivalence of $Z({\rm Vec}_\cG)$, as it exchanges a boson with a fermion.

Second, consider $\cG=\ZZ_2\times\ZZ_4$. In addition to the eight pure fluxes, the bosons $m_1e_2^2$ and $m_2e_1$ appear in an SPT phase: the phase represented by the cocycle $\omega(g,h)=\tfrac{1}{2}g_1h_2$ realizes $m_1e_2^2$ as the slant product $(g,i_g\omega)$ \eqref{eq:SPT-slant} for $g=(1,0)$ and $m_2e_1$ for $g=(0,1)$. The non-charge bosons $m_1e_2$, $m_1e_2^3$ and $m_1^2e_2$, on the other hand, do not appear in any SPT phase and so have the potential to be exchanged with charges under an SPT-stabilizing duality. In fact the ``dual stacking'' duality -- given by EM duality $e_i\leftrightarrow m_i$, composed with SPT stacking by $\omega$ ($m_1e\rightarrow m_1e_2^2e$ and $m_2e\rightarrow m_2e_1e$, for any charge $e$), composed with EM duality again -- does exactly this: it exchanges $m_1e_2$ with $e_2$,  $m_1e_2^3$ with $e_2^3$, and $m_1^2e_2$ with $e_1$. But despite not stabilizing $\Rep(\cG)$, this duality acts on SPT phases like one that does \eqref{eq:repGstab-action}; namely, by fixing the pure fluxes and exchanging the bosons $m_1e_2^2$ and $m_2e_1$, it acts trivially on the set of SPT phases.\footnote{For an example of a duality that acts nontrivially on the set of SPT phases (but still according to the rule \eqref{eq:repGstab-action}), consider dual stacking by $\omega$ composed with regular stacking by $\omega$.}

Third, consider $\cG=\ZZ_8\times\ZZ_8$. This time, dual stacking by the generating phase $\omega(g,h)=\tfrac{1}{8}g_1h_2$ is not a duality of SPT phases, as it takes $me_1^ie_2^j$, for any flux $m$, to $mm_1^{-j}m_2^ie_1^ie_2^j$, so in particular $m_2e_1^{-1}$ (a boson of the SPT phase $\omega$) maps to the pure charge $e_1^{-1}$ (a boson in no SPT phase). Dual stacking by $2\omega$, on the other hand, \emph{does} map between SPT phases: the generating bosons $m_1e_2^k$ and $m_2e_1^{-k}$ of the SPT phase $k\omega$ map to the bosons $m_1^{1-2k}e_2^k$ and $m_2^{1-2k}e_1^{-k}$, which taken to the power $1-2k$ (and using $(1-2k)(1-2k)\equiv 1\,{\rm mod}\,8$) yields $m_1e_2^{k(1-2k)}$ and $m_2e_1^{-k(1-2k)}$; thus, this duality acts on the $\ZZ_8$ group of SPT phases according to
\begin{equation}
    k \mapsto k(1-2k)=\left\{\begin{array}{ll}k~,&k\,{\rm even}\\k+6~,&k\,{\rm odd}\end{array}\right.~.
\end{equation}
Notably, this duality fixes the trivial phase yet does not act the SPT classification by a group automorphism; it is not of the form \eqref{eq:repGstab-action}. This exotic action raises the question of whether the action $\cJ\cW^{-1}\cX$ \eqref{jwinvX} related to the stacking isomorphism can be implemented by a duality.

Let us continue to consider $\cG=\ZZ_8\times\ZZ_8$. Dual stacking by $4\omega$ performs twice the action of dual stacking by $2\omega$, so it shifts odd phases by $4$. This is equivalent to multiplying all phases by $5$, which is an automorphism. Realize the group as a fermionic symmetry $\cG=\ZZ_8\times\ZZ_8^f$ by taking $p=(0,4)$ to be fermionic parity. The phase $k\omega$ corresponds to supercohomology variables $(k\alpha_0,(k\,{\rm mod}\,2)\beta_0)$. In these terms, dual stacking by $4\omega$ takes the form $\cJ\cW^{-1}\cX:[\alpha,\beta]\mapsto[\alpha-\Lambda(\beta),\beta]$ \eqref{jwinvX} with $\Lambda(\beta_0)=4\alpha_0$. Note that $\delta\Lambda=0$, which is forced by $\cJ\cW^{-1}\cX$ being an automorphism and $\cJ\cW$ ($\Lambda=0$) being an isomorphism $\cB\rightarrow\cF$. (For a simpler example, consider dual stacking by $2\omega$ on $\ZZ_4\times\ZZ_4$, which has $\Lambda(\beta_0)=2\alpha_0$.)

Fourth, consider $\cG=\ZZ_4\times\ZZ_4\times\ZZ_4^f$. Since $\beta_1,\beta_2\in\BB$ (on the first two factors) have nonvanishing half cup product in $\AAA$, $\cJ\cW$ is not an isomorphism here. But there exist stacking group isomorphisms given by the $\Lambda$ that trivialize $\Omega_\cF$, and we can ask whether these are implemented by dualities. Since $\cJ\cW^{-1}\cX$ is not an automorphism, we might try to cook up a dual stacking duality similar to the previous example. For $\ZZ_4$ factors $i\ne j=1,2,f$, write $\omega_{ij}(g,h)=\tfrac{1}{4}g_ih_j$. By using calculations like the above, one can show that dual stacking by $2\omega_{12}$ maps $\omega_{12}$ to $3\omega_{12}$ while fixing $\omega_{1f}$ and $\omega_{2f}$; it also maps $\omega_{12}+\omega_{1f}$ to $3\omega_{12}+3\omega_{1f}$, which means it fails to be an automorphism, as desired. To get a map of the form $\cJ\cW^{-1}\cX:[\alpha,\beta]\mapsto[\alpha-\Lambda(\beta),\beta]$, we need to fix phases of the form $[\alpha,0]$; that is, phases generated by $\omega_{12}$, $2\omega_{1f}$, and $2\omega_{2f}$. This is accomplished by the duality given by dual stacking with $2\omega_{1f}+2\omega_{2f}$, which acts on $a\omega_{12}+b\omega_{1f}+c\omega_{2f}$ as
\begin{equation}
    (a,b,c)\mapsto ((-1)^{b+c}a,(-1)^{c+1}b,(-1)^{b+1}c)~.
\end{equation}
But this transformation does not have the correct form, as the shift it performs on $\alpha$ is not a function $\Lambda(\beta)$ but rather depends on $\alpha$ as well. As we will see in the general proof below, it turns out to be impossible to realize any of this group's stacking isomorphisms by a duality.

Fifth, consider $\cG=\ZZ_2\times\ZZ_2\times\ZZ_2^f$. The automorphism of $\cG$ that maps the cyclic generators to $(1,0,0)$, $(0,1,1)$, and $(0,0,1)$ induces a duality between SPT phases. The fluxes $m_1, m_2, m_f$ map to $m_1, m_2m_f, m_f$; the charges $e_1, e_2, e_f$ must map to $e_1, e_2, e_2e_f$ to preserve braiding. Consider the action on the phase $\omega_{1f}$: its anyons $m_1e_f, m_2, m_fe_1$ map to $m_1e_2e_f, m_2m_f, m_fe_1$, which generate the same algebra as $m_1e_2e_f$, $m_2e_1$, $m_fe_1$, which correspond to the phase $\omega_{12}+\omega_{1f}$. Similarly, one can compute how the other phases transform: for example, $\omega_{12}$ and $\omega_{2f}$ are fixed, while $\omega_{1f}+\omega_{2f}$ maps to $\omega_{12}+\omega_{1f}+\omega_{2f}$. It turns out that this duality has the form $[\alpha,\beta]\mapsto[\alpha-\Lambda(\beta),\beta]$ with $\Lambda(\beta_1)=\omega_{12}=\tfrac{1}{2}\beta_1\cup\beta_2$, $\Lambda(\beta_2)=0$, and $\Lambda(\beta_1+\beta_2)=\omega_{12}$. Note, however, that even though the desired $\omega_{12}$ shift appears in $\Lambda$, we still have $\delta\Lambda=0\ne\Omega_\cF$, so this is not a stacking isomorphism. (This is no surprise, as we found in \cref{dualities:example} that no duality implements the stacking isomorphism for this symmetry.) In general, an automorphism $a$ of $\cG$ induces a duality that takes the SPT phase $\omega$ to the phase $a^*\omega$ since each $(g,i_g\omega)$ maps to $(a^{-1}(g),a^*i_g\omega)=(a^{-1}(g),i_{a^{-1}(g)}(a^*\omega))$. The action $a^*$ is an automorphism of SPT phases, so it can only ever be a stacking isomorphism if $\delta\Lambda=0$.

With these examples in mind, we are ready to prove the theorem of \cref{sec:dualities_Lag_SPT}.

First, let us see that every duality that fixes the trivial SPT phase consists of a dual stacking and an automorphism of $\cG$ (the order is irrelevant since the inverse duality also fixes the trivial SPT phase). Fixing the trivial phase means mapping pure magnetic fluxes to pure magnetic fluxes, so this statement is equivalent to the dual statement \eqref{eq:repGstab-action} (due to Ref. \cite{nikshych2013categorical}) about the charges $\Rep(\cG)$. For completeness, let us present a self-contained proof of this fact. Suppose a duality $\cD$ maps charges to charges: $\cD(1,\chi)=(1,\chi')$. Since $\cD$ is compatible with fusion, $\chi'=a^*\chi$ for some automorphism $a:\cG\ra\cG$. On fluxes, write $\cD(g,1)=(g',f(g'))$. Then $\cD(g,\chi)=\cD(g,1)\cdot\cD(1,\chi)=(g',f(g'))\cdot(1,a^*\chi)=(g',f(g')a^*\chi)$. Since $\cD$ is braided, it preserves topological spins, so $f(g')(g')=1(g)=1$ and so $\chi(g)=(f(g')a^*\chi)(g')=(a^*\chi)(g')$; thus, $g'=a^{-1}(g)$. Since $\cD$ is compatible with fusion, $f:\cG\ra\cG^*$ is a homomorphism. By evaluating its image on $\cG$, $f$ defines a bicharacter $\cG\times\cG\ra U(1)$. This bicharacter is antisymmmetric in its arguments since $f(g)(g)=\theta(g,f(g))=\theta(\cD(a(g),1))=\theta(a(g),1)=1$, where $\theta$ denotes topological spin. All antisymmetric bicharacters arise from slant products of cocycles $\zeta$ as $f(g)=i_g\zeta$ (c.f. Ref. \cite{berkovich2017yakov}, Def. VI.6.1 and Theorem VI.6.3). Thus we conclude that $\cD(g,\chi)=(a^{-1}(g),i_{a^{-1}(g)}\zeta\cdot a^*\chi)$. The bosons of the SPT phase $\omega$ are mapped to $\cD(g,i_g\omega)=(a^{-1}(g),i_{a^{-1}(g)}\zeta\cdot a^*(i_g\omega))=(a^{-1}(g),i_{a^{-1}(g)}(\zeta+a^*\omega))$, which belong to the phase $\zeta+a^*\omega$. This completes the proof of Eq. \eqref{eq:repGstab-action}. Note that, in the dual version, not all dual stackings map SPT phases to SPT phases; the point is that any duality of SPT phases that fixes the trivial phase is realized this way.

Next, we prove the theorem by arguing that dual stackings and $\cG$ automophorisms cannot act on SPT phases as $\cJ\cW^{-1}\cX:[\alpha,\beta]\mapsto[\alpha-\Lambda(\beta),\beta]$ for $\delta\Lambda=\Omega_\cF$, unless $\cJ\cW$ is an isomorphism ($\Omega_\cF=0$). Suppose $\cJ\cW$ is not an isomorphism, which according to the discussion in \cref{sec:groupexts} means $\cG$ is of the form $\cG=H\times\ZZ_{n_1}\times\ZZ_{n_2}\times\ZZ_{n_f}^f$ for $n_{if}:=\gcd(n_i,n_f)=n_f$ for $i=1,2$. Consider the valid $\beta_i$ that evaluate to $1$ on the generators of $\ZZ_{n_i}$ and zero on the others. Their half cup product is $\Omega_\cF(\beta_1,\beta_2)=\tfrac{n_{12}}{2}\omega_{12}$. It will turn out that any dual stacking between SPT phases acts on $\omega_{1f}$, $\omega_{2f}$, and $\omega_{1f}+\omega_{2f}$ as
\begin{equation}\label{dstack-act}
    \omega_{if}\mapsto k_i\omega_{if}~,\qquad\omega_{1f}+\omega_{2f}\mapsto k_3(\omega_{1f}+\omega_{2f})~,
\end{equation}
for $k_1,k_2,k_3$ odd. We will prove this below. For now, let us use it to prove the theorem.

Suppose the duality $\cD$ acts on SPT phases like $[\alpha,\beta]\mapsto[\alpha-\Lambda(\beta),\beta]$ for some $\Lambda$. Then $\delta\Lambda$ is given by
\begin{align}\begin{split}\label{dLfromduality}
    \delta\Lambda(\beta_1,\beta_2)&=\cD(\omega_{1f})+\cD(\omega_{2f})-\cD(\omega_{1f}+\omega_{2f})\\
    &=a^*(k_1\omega_{1f})+a^*(k_2\omega_{2f})-a^*(k_3(\omega_{1f}+\omega_{2f}))\\
    &=xa^*(\omega_{1f})+ya^*(\omega_{2f})~,
\end{split}\end{align}
for $x=k_1-k_3$, $y=k_2-k_3$, using the lemma \eqref{dstack-act} and the fact that $a^*$ is an automorphism. Let us see that $\delta\Lambda(\beta_1,\beta_2)=\tfrac{n_{12}}{2}\omega_{12}$ is impossible. Since this class is of order $2$, we must have $x,y\in\{0,\tfrac{n_f}{2}\}$. The rule \eqref{dstack-act} means the dual stacking does not alter the $\beta$ part of a phase; thus, since $\cD$ does not alter $\beta$, then $a^*$ must not either. Therefore $a^*$ must map $\omega_{if}$ to an odd multiple of itself plus other terms not involving any $\omega_{jf}$. Therefore, the right-hand side $\tfrac{n_f}{2}a^*(\omega_{1f})+\tfrac{n_f}{2}a^*(\omega_{2f})$ either vanishes or still contains a multiple of $\omega_{1f}$ or $\omega_{2f}$. We conclude that it cannot equal $\Omega_\cF(\beta_1,\beta_2)=\tfrac{n_{12}}{2}\omega_{12}$, thus proving the theorem.

It remains to show that dual stackings between SPT phases act according to the rule \eqref{dstack-act}. Consider the phase $\omega_{1f}$. It has dyons $m_1e_f$ and $m_fe_1^r$ (where $r=-n_1/n_f$) as well as pure fluxes $m_i$ for $i\ne 1,f$. Dual stacking fixes pure fluxes, so, assuming it maps $\omega_{1f}$ to some other SPT phase, it must map it to a phase with the same fluxes, i.e. to a multiple $k_1\omega_{1f}$. A dual stacking takes the dyon $m_1e_f$ of $\omega_{1f}$ to one of the form $m_1^{z+1}(m_i\cdots)e_f$. This dyon belongs to $k_1\omega_{1f}$, so $e_f=i_{m_1^{z+1}(m_i\cdots)}(k_1\omega_{1f})=(i_{m_1}\omega_{1f})^{(z+1)k_1}=e_f^{(z+1)k_1}$. Thus $k_1$ is odd. A similar argument about the phase $\omega_{2f}$ shows that $k_2$ is odd.

Now we turn to $\omega_{1f}+\omega_{2f}$, which has dyons $m_1e_f$, $m_2e_f$, and $m_fe_1^re_2^s$ and pure fluxes $m_i$ for $i\ne 1,2,f$. Since pure fluxes are fixed by dual stackings, the resulting phase has the form $q\omega_{12}+x\omega_{1f}+y\omega_{2f}$. First we will show that $q=0$. Dual stacking takes the pure charge $e_1$ to the dyon $m_f^z(m_i\cdots )e_1$. If $z$ is odd, it has a modulo $n_f$ multiplicative inverse $\bar z$, and taking the dyon to this power is $m_f(m_i\cdots)e_1^{\bar z}$, which appears in the SPT phase $\omega_{1f}^{\bar z}$. Since a duality between SPT phases never maps a pure charge (contained in no SPT phase) to a boson belonging to an SPT phase, $z$ must be even. The dual stacking acts on the dyons of $\omega_{1f}+\omega_{2f}$ by taking $m_1e_f$ to $m_1^{z+1}(m_i\cdots)e_f$, $m_2e_f$ to $m_2^{z'+1}(m_i\cdots)e_f$, and $m_fe_1^re_2^s$ to $m_f^u(m_i\cdots )e_1^re_2^s$, where $u=1-zr-z's$. Since $z$ and $z'$ are even, $u$ is odd, which means that all nonvanishing powers of $m_f^u(m_i\cdots )e_1^re_2^s$ contain a factor of $m_f$. Therefore, the only dyons of the resulting phase involving an $e_1$ or $e_2$ charge involve a multiple of $m_f$, so there is no $\omega_{12}$ term, i.e. $q=0$. Finally, let us show that $x$ and $y$ are odd. The dual stacking takes $m_1e_f$ to $m_1^{z+1}m_2^{z'}(m_i\cdots)e_f$, so $e_f=i_{m_1^{z+1}m_2^{z'}(m_i\cdots)}(x\omega_{1f}+y\omega_{2f})=(i_{m_1}\omega_{1f})^{(z+1)x}(i_{m_2}\omega_{2f})^{z'y}=e_f^{(z+1)x+z'y}$, so $(z+1)x+z'y=1$. Similarly, $m_2e_f$ maps to $m_1^zm_2^{z'+1}(m_i\cdots)e_f$, so $zx+(z'+1)y=1$. Together, these imply $x=y$. Then $1=(z+1)x+z'y=(z+z'+1)x$ implies that $x=y$ is odd. This proves the lemma \eqref{dstack-act}.

\printbibliography

\end{document}